\RequirePackage{rotating}
\documentclass[a4paper,fleqn,usenatbib]{mnras}

\usepackage[T1]{fontenc}
\usepackage{ae,aecompl}
\usepackage{float}

\usepackage{graphicx}	% Including figure files
\usepackage{amsmath}	% Advanced maths commands
\usepackage{amssymb}	% Extra maths symbols
\usepackage{txfonts}
\usepackage{ulem}
\usepackage{xcolor}
\newcommand{\Figs}[1]{Figs.~\ref{#1}}
\newcommand{\Sec}[1]{\S\ref{#1}}

\newcommand{\gpk}{\hbox{\sc GalPaK$^{\rm 3D}$}}{}
{}
\newcommand{\papone}{paper\,I}{}
\newcommand{\paptwo}{paper\,II}{}
\newcommand{\papthree}{paper\,III}{}
{}

\title[The CGM dust-anisotropy]{MusE GAs FLOw and Wind (MEGAFLOW) V. The dust/metallicity-anisotropy of the Circum-Galactic Medium~
\thanks{Based on observations made at the ESO telescopes at La Silla Paranal Observatory under programme IDs
094.A-0211,
095.A-0365,
096.A-0164,
096.A-0609, %UVES
097.A-0138,
097.A-0144, %UVES
097.A-0144,
098.A-0216,
098.A-0310, %UVES
099.A-0059,
293.A-5038,
0100.A-0089,
0101.A-0287, and/or data products created thereof.}} %UVES,
\author[Wendt et al.]{
M. Wendt,$^{1,2}$\thanks{E-mail: mwendt@astro.physik.uni-potsdam.de}
Nicolas F. Bouch\'e,$^{3}$
Johannes Zabl,$^{3}$
Ilane Schroetter,$^{4}$
Sowgat Muzahid$^{5,2}$
\\
$^1$ Institut f\"ur Physik und Astronomie, Universit\"at Potsdam,Karl-Liebknecht-Str. 24/25,  
14476 Golm, Germany\\
$^2$ Leibniz-Institut f\"ur Astrophysik Potsdam (AIP), An der Sternwarte 16, 14482 Potsdam, Germany \\
$^3$ Univ Lyon, Univ Lyon1, Ens de Lyon, CNRS, Centre de Recherche Astrophysique de Lyon UMR5574, F-69230 Saint-Genis-Laval, France\\
$^4$ GEPI, Observatoire de Paris, CNRS-UMR8111, PSL Research University, Univ. Paris Diderot, 5 place Jules Janssen, 92195 Meudon, France \\
$^5$ IUCAA, Post Bag 04, Ganeshkhind, Pune-411007, India\\
}

\date{Received \today; accepted \today}
\pubyear{2021}

\newcommand{\MgI}{\hbox{{\rm Mg}{\sc \,i}}}
\newcommand{\HI}{\hbox{{\rm H}{\sc \,i}}}

\newcommand{\MgII}{\hbox{{\rm Mg}{\sc \,ii}}}
\newcommand{\FeII}{\hbox{{\rm Fe}{\sc \,ii}}}
\newcommand{\CaII}{\hbox{{\rm Ca}{\sc \,ii}}}
\newcommand{\CrII}{\hbox{{\rm Cr}{\sc \,ii}}}
\newcommand{\ZnII}{\hbox{{\rm Zn}{\sc \,ii}}}
\newcommand{\MnII}{\hbox{{\rm Mn}{\sc \,ii}}}
\newcommand{\TiII}{\hbox{{\rm Ti}{\sc \,ii}}}
\newcommand{\OII}{\hbox{{\rm O}{\sc \,ii}}}
\newcommand{\km}        {\ensuremath{\mathrm{\ km}}}
\newcommand{\s}         {\ensuremath{\mathrm{\ s}}}
\newcommand{\kms}       {\ensuremath{\mathrm{\km\s^{-1}}}}

\newcommand{\Ly}{\hbox{{\rm Ly}$\alpha$}}

\begin{document}
\label{firstpage}
\pagerange{\pageref{firstpage}--\pageref{lastpage}}
\maketitle
\begin{abstract}
We investigate whether the dust content of the circum-galactic medium (CGM) depends on the location of the quasar sightline with respect to the galaxy major-axis
using 13 galaxy-\MgII{} absorber pairs (9 - 81 kpc distance) from the MusE GAs FLOw and Wind (MEGAFLOW) survey
at $0.4<z<1.4$. The dust content of the CGM is obtained from [Zn/Fe] using  Ultraviolet and Visual Echelle Spectrograph (UVES) data.
When a direct measurement of [Zn/Fe] is unavailable, we  estimate the dust depletion from a method which consists in solving for the depletion from multiple singly ionized ions (e.g. \MnII{},  \CrII{}, \ZnII{}) since  each ion depletes on dust grains at different rates.
We find a positive correlation between the azimuthal angle and [Zn/Fe] with a Pearson's $r = 0.70 \pm 0.14$. The sightlines along the major axis show [Zn/Fe] < 0.5, whereas the [Zn/Fe] is > 0.8 along the minor axis. 
These results  suggest that the CGM along the minor
axis is on average more metal enriched (by $\approx$ 1 dex) than the gas located along the major axis
of galaxies provided that dust depletion is a proxy for metallicity.
This anisotropic distribution is  consistent with recent results on outflow and accretion in hydro-dynamical simulations.
\end{abstract}


%%%%%%%%%%%%%%%%%%%%%%%%%%%%%%%%%%%%%%%%%%%%%%%%%%

%%%%%%%%%%%%%%%%% BODY OF PAPER %%%%%%%%%%%%%%%%%%

\section{Introduction}
In the $\Lambda$CDM cosmological context,  supernovae(SN)-driven outflows \citep{dekelsilk86} are thought to play a major role in galaxy formation
given that galactic winds are ubiquitous and  necessary to account for the observed mass-metallicity relation. They could eject a significant fraction of the baryons and play  a key role in enriching the inter-galactic medium \citep{Veilleux2005, Veilleux:2020a, Heckman:2017}.
 SN-driven outflows and gas inflows usually have rather low particle densities
 and are hence difficult to observe.
 The properties of this low density gas can, however, be studied through quasar absorption line systems (QSOALs). The circum-galactic medium (CGM)
 is typically extended out to a few 100 kpc around galaxies. The physical and chemical conditions of the CGM are determined by gas infall, outflow, and other dynamical processes. 
Over the past decades, 
various surveys have been developed to study the distribution, physical state, and kinematics of the  CGM around galaxies 
using both metal absorption lines
(e.g., \citealp{Steidel:2010, Chen:2010, Prochaska:2011, Bordoloi:2011a,Werk:2013a,Nielsen:2013a,Schroetter:2016a,  Rubin:2018a, Zabl:2019a, muzahid2019} and references therein). 
 
Galaxy-absorber pairs at high redshifts ($2.5<z<2.9$) were studied by \citet{Rudie:2012}.
Therein 15 QSO sightlines from the Keck Baryonic Structure Survey (KBSS) were analyzed
 with the strongest \HI\ absorbers within $\approx$ 100 kpc of the nearest galaxy.
At intermediate redshifts, when \Ly\ in the ultraviolet (UV) is not available from the ground, the CGM can be best studied with metal lines
in quasar spectra. While, it is well established since the early 90s \citep{Bergeron:1991a, Bergeron:1992, Steidel:1992}, that strong \MgII\ absorbers with rest-frame $W^{\lambda2796}_{\rm r} > 0.3$\AA\, are known to be generally associated with $\approx L_\star$ star-forming galaxies. The resonant $\MgII{}\,\lambda\lambda 2796, 2803$ doublet is an especially useful tracer of the cool ($10^{4\mbox{--}5}\,\mathrm{K}$), metal enriched CGM due to its strength and rest-wavelength, which allows to study it with ground-based optical spectroscopy for redshifts between $0.3\lesssim z \lesssim 2.5$. 

Among galaxy-absorber pair \MgII{} surveys, there is the Keck survey of Martin and collaborators \citep[e.g.][]{Ho:2017a,Ho:2019b} at lower redshifts ($z\sim0.3$), who find that the 
broad velocity shifts of the \MgII{} absorption can successfully be explained by gas spiraling inwards near the disk plane, as found by \citet{Bouche:2013a,Bouche:2016a} and \citet{Zabl:2019a}.
One of the largest surveys of \MgII{} absorbers is the \MgII{} Absorber-Galaxy Catalog (MAGIICAT) which contains 182 galaxies with QSO sightlines within 200 kpc \citep{Nielsen:2013a}.
 \citet{Rubin:2018a}  built a sample of 27 \MgII\ absorber-QSO pairs with galaxies in the PRIMUS survey and background QSO as well as other galaxies   in the redshift range of $0.4<z<1.0$.  
Earlier, \citet{Churchill:1999}  observed 30 weak ($W_r^{\lambda2796}$ < 0.3 \AA) \MgII{} absorbers at a similar redshift 
range ($0.4<z<1.4$) and associate those weak \MgII\ absorbers with sub Lyman limit system (LLS) environments \citep[see also][]{Muzahid:2018a}.
A study on a possible dust selection  bias for samples selected for strong \MgII\  absorbers ($W_r^{\lambda2796}$ > 0.5 \AA)
in the Complete Optical and Radio Absorption Line System (CORALS) survey at redshifts $0.7<z<1.6$ finds
no particularly high dust-to-metals ratios in their sample \citep{Ellison:2009}.
 
Thanks to these surveys and others, there has been an important advance of our understanding of  the statistics of the CGM, namely on the anisotropic distribution of the presence of \MgII{} absorption lines since \citet{Bordoloi:2011a} and
\citet{Bouche:2012a}. This anisotropic presence of \MgII{} around galaxies pointed towards a dual physical mechanism at play, namely that \MgII{} QSOALs are caused either by the accreting/rotating extended gaseous disks or by outflows \citep[see also][]{Lan:2018a}.
This scenario is now strongly supported from the relation between the CGM kinematics and the host-galaxy kinematics   \citep[e.g.][]{Bouche:2013a,Bouche:2016a,Ho:2017a,Ho:2019b,Kacprzak:2014a,Martin:2019a,Muzahid:2015a,Schroetter:2016a,Schroetter:2019a,Zabl:2019a, Rahmani:2018a, Rahmani:2018b}.

In this setting, the gas in the SN-driven outflows ought to be more dusty (as in M82) or metal enriched than the co-planar material (presumably related to accretion).  
However, while there is strong evidence that the CGM gas metallicities are bimodal for LLs at $0.45 < z < 1.0$
\citep{Lehner:2013a,Lehner:2016a,Wotta:2016,Wotta:2019},  there is currently no supporting observational evidence that the anisotropic bimodal gas distribution for \MgII{} absorbers is related to the bimodal metallicity distribution for LLS \citep{Pointon2019}.  Furthermore, comparing the CGM metallicity to the metallicity of the interstellar medium (ISM) of the galaxy shows that gas flow metallicities   have a wide range of  values  compared  to  that of the  host  galaxies  regardless  of the azimuthal angle  \citep{Peroux:2016a,Peroux:2017a,Kacprzak:2019a}. 
It should be noted, in regards to these null results, that  the \HI\ selection of \citet{Peroux:2017a} appears to select complex CGM situations when there are multiple galaxies within 100-150kpc. In addition, the sample of  \citet{Pointon2019} is likely  not uniformly selected given that their survey relies on photometric pre-selections of follow-up galaxies, a limitation that can be avoided with integral field unit (IFU) spectroscopy surveys.

In this context, we aim to use our  MUSE GAs FLOw and
Wind (MEGAFLOW) survey \citep{Schroetter:2016a, Schroetter:2019a, Zabl:2019a, Zabl:2019b} (paper~I, III, II, IV)
to investigate dust (and metallicity) properties of the CGM with respect to the anisotropic \MgII{} distributions.
MEGAFLOW, which aimed   at studying the kinematics and physical properties of $\sim$ 100 star-forming galaxies at $0.4<z<1.4$ around   \MgII{} absorptions with rest-frame equivalent width ($W_r^{\lambda2796}$) larger than 0.5\AA, has enabled us  to investigate the anisotropic distribution of the CGM gas traced by  \MgII{} absorption lines \citep[see][]{Schroetter:2019a,Zabl:2019a}. 
MUSE observations have been used in the past to analyze individual projected quasar galaxy pairs with regard to 
the CGM-host galaxy connection \citep[e.g.,][]{Bouche:2016a, Rahmani:2018b}.

In this paper, we use the large sample of galaxy-\MgII{} absorber pairs from the MEGAFLOW survey which already showed strong evidence for \MgII{} anisotropy \citep{Schroetter:2019a,Zabl:2019a}  in order
to study whether also the amount of CGM dust-depletion depends on the alignment between absorber and galaxy.
Given that our survey redshift range does not allow us to measure metallicities directly (as \HI\ falls in the far ultra-violet), we will study in this paper the dust content of our sightlines using the prescription of \citet{DeCia:2016a}.

The paper is organized as follows: we present our sample in \Sec{sec:data},  the methodology in \Sec{sec:methodo}, and the results  in \Sec{sec:results}. Finally, we discuss our results and we present our conclusions  in \Sec{sec:conclusions}.
Throughout the paper, we use a 737 cosmology ($H_0=70$~\kms Mpc$^{-1}$, $\Omega_{\rm m}=0.3$, and $\Omega_{\Lambda} = 0.7$).

\section{Data}
\label{sec:data}
We use the data from the MEGAFLOW survey \citep[hereafter \papone]{Schroetter:2016a} that was published in \citet[\paptwo]{Zabl:2019a} and \citet[\papthree]{Schroetter:2019a}.
This survey has enabled us to bring the sample size of galaxy-quasar pairs with robust galaxy orientation information from a dozen \citep{Bouche:2012a, Schroetter:2015a} to 79 by targeting 22 quasar fields  where { each sightline has {\it multiple} (3 to 5) \MgII{} absorbers} at redshifts accessible with [\OII{}] in MUSE at $0.4<z<1.4$, taking full advantage of the multiplexing capabilities of this 3D spectrograph \citep{Bacon:2010a}. 
Paper\,III covers the galaxies with corresponding absorption of the winds, while  \paptwo{} lists cases attributed to accretion.

The division into accretion and wind cases was made based on the azimuthal angle.
The azimuthal angle, $\alpha$, is defined as the angle between the location of the background quasar on the sky and the (projected) major axis of the galaxy associated to the absorber (see e.g. Figure 1 of \paptwo{} for an illustration). Quasar sightlines which are close to the galaxy's minor axis ($\alpha=90\deg$) are likely to probe outflows (\papthree), while sightlines along the major axis ($\alpha=0\deg$) are passing through extended gas disks (\paptwo).
In practice, we determined the major-axis by fitting a 3D morpho-kinematical model to the [\OII{}] emission doublet using the 3D fitting tool \gpk{} \citep{Bouche:2015a}.

For the MEGAFLOW survey, each MUSE field was searched
for [\OII{}] in emission corresponding to the \MgII{} absorption redshifts
seen in the quasar spectrum within a velocity interval
 of $\pm$1\,000 \kms. The details of this process are described in \papone.

For each galaxy-absorber pair in the full MEGAFLOW sample, we searched for absorption
of weaker ions in high resolution VLT/UVES data (dedicated observations, see section \ref{sec:uves}) to derive the depletion via the model given in \citet{DeCia:2016a} and selected those.
Independently, all pairs for which we determined a robust measurement of the azimuthal angle from the 3D morpho-kinematical model
were considered (see \paptwo{} and \papthree). Merging these two selection criteria brought down the 
sample to 13 galaxy-absorber pairs for which we detected singly ionized elements that enabled us to apply
the prescription in \citealp{DeCia:2016a} (see section \ref{sec:methodo}). For those 13 pairs, whenever feasible, we measured the dust depletion [Zn/Fe] directly via Zn which is minimally affected by depletion and Fe with possible depletion onto dust grains (4 cases, see Table~\ref{table:targets}).
The precise redshifts were already determined by the dominant \MgII{} absorption.
When information of the complex velocity structures was available from stronger elements such as 
\FeII{}, we felt confident identifying weak ions in single transitions for individual cases.

\section{UVES Observations and data reduction}
\label{sec:uves}
The quasar fields of our MUSE GTO-Programme were observed with the high-resolution spectrograph UVES \citep{Dekker:2000a} between 2014 and 2016 (Table~\ref{table:targets}).
The settings used in our observation were chosen in order to cover the \MgII$\lambda \lambda 2796,2803$ absorption lines as well as other elements such as \MgI$\lambda 2852$, \FeII$\lambda 2586$ when possible. The full list of observations is given in \paptwo{} and \papthree.

The data were taken under similar conditions resulting in a spectral resolving power of R $\approx 38000$ dispersed on pixels of $\approx$1.3 \kms. 
The Common Pipeline Language (CPL version 6.3) of the UVES pipeline was used to bias correct and flat field the exposures and then to extract the wavelength and flux calibrated spectra. 
After the standard reduction, the custom software UVES Popler \citep[][version 0.66]{MurphyM_16a}  
was used to combine the extracted echelle orders into single 1D spectra. 
The continuum was fitted with low-order polynomial functions. 

For this analysis we searched for the singly ionized species to determine the depletion of the absorption system such as the \MnII{}, \ZnII{}, \CrII{} and \TiII{}.
Of the 45 
galaxy-absorber pairs in our MEGAFLOW sample, we were able to identify and fit these ions in 13 pairs.
All absorption features were modeled as Gaussians with the evolutionary algorithm described in \cite{Quast2005}
 that is particularly successful in deblending complex velocity structures \citep{Wendt2011}.
The same velocity components were fitted --when detected-- for every ion.  All observed transitions of an individual ion were fitted with a common broadening parameter and column density and the fitted transitions are shown in the Appendix (see Fig.\ref{fig:depletion:examples}). As a consequence, the column densities of \FeII{} are rather precise thanks to the utilization of weak transitions such as \FeII{} 2586 and 2374 with oscillator strengths as low as 0.069 and 0.031, respectively \citep[see][]{Cardelli1995} that constrain the overall column density of \FeII{}.
In some cases the weaker ions could not be deblended into distinct velocity components and as all our singly ionized elements are optically thin, we additionally derived logN from equivalent width measurements via linear relation for those systems via:
$$N(\mathrm{cm}^{-2}) = 1.13 \times 10^{20} W_\lambda   f_{\mathrm{osc}}^{-1}\lambda^{-2}$$ with $W_\lambda$ and $\lambda$ in units of \AA.
 The column densities for the ions used in this paper are listed in Table \ref{table:columndensities}.
 
Finally, to evaluate how representative our selection of the 13 galaxy-absorber pairs is with regard to the overall MEGAFLOW sample and the
45 galaxy-absorber pairs therein, we compared their distributions of impact parameters.
The corresponding two-sided KS tests of the distribution for the 45-13 = 32 absorbers not part of this paper and the 13 absorbers used in this work has a p-values of 0.18. We thus conclude that the 13 absorbers with regard to impact parameters are typical of the MEGAFLOW sample.

Our sample based on the detection of weaker ions favors \MgII{}\,systems with slightly higher restframe EWs on average compared to the full sample (median value of 
1.86\AA{} compared to the median EW of 1.25\AA{} for all 79 \MgII{}\,systems).
The two-sided KS test yields that the hypothesis that they follow the same distribution in EWs can still not be rejected beyond the 2-$\sigma$ level.
For comparison to other samples, it is noteworthy that the \citet{Pointon2019} sample has a quite different impact parameter distribution and that, in most cases,
their systems are no strong \MgII{} absorbers. 

\section{Methodology}
\label{sec:methodo}
 Given the redshift range of our systems ($0.4<z<1.4$) imposed by the MUSE wavelength coverage of [\OII], a direct metallicity $[X/H$] estimation is currently difficult since the corresponding Lyman series lines fall in the near UV (NUV) and are often inaccessible even with HST/COS given the low  NUV fluxes of our QSOs.

Here, we aim at measuring the dust depletion in the CGM as a function of azimuthal angle.
This can be easily achieved because several authors have shown that 
 various elements are depleted at different rates \citep[e.g.][]{Savage:1996a,Vladilo:2002a,Vladilo:2002b,Jenkins:2009a,DeCia:2016a} where Zn is the least depleted element \citep[e.g.][]{Vladilo:2000a}. 
In particular, \citet{Jenkins:2009a} built a framework that describes the depletion in terms of a set of a simple parameter, $F_\star$, the depletion strength factor, using data from sightlines of Milky-Way halo stars, i.e. at around solar metallicity.
More recently,  \citet{DeCia:2016a} extended the method of \citet{Jenkins:2009a} to intergalactic clouds (colum density N$_{\HI}$ > $10^{20}$ cm$^{-2}$) and showed that relative abundances of  intergalactic and galactic sightlines can all be put into a similar coherent framework\footnote{One limitation of the \citet{Jenkins:2009a} framework is that $F_\star$ was calibrated in the MilkyWay enriched environment such that the lowest depletion $F_\star=0$ correspond to  [Zn/Fe]=1 whereas  this quantity can range below 1 in DLAs.} where they parameterized the depletion sequence of each element as a function of  [Zn/Fe].\footnote{See also \citet{Jenkins:2017a} and \citet{Wiseman:2017a} for other extensions of the original \citet{Jenkins:2009a} framework to  GRBs and the SMC, respectively.}

The methodology can be understood as follows:
given that the depletion [X/Zn] of an element $X$ varies linearly with [Zn/Fe] with each element having  a different slope ($B_2$ in \citealp{DeCia:2016a}), the global depletion factor    [Zn/Fe]  can be determined if one has  at least two ions, even when a direct Zn measurement is not available by solving a simple set of linear equations.  The model fits are shown in Fig.~\ref{fig:depletion:examples}. The two unknowns are the global depletion [Zn/Fe] given by the slope and $\log N_{\mathrm{H}}+\log Z$ given by the zero-point  of the line when each element abundance
is plotted  as a function of the  depletion propensity  $B_2$.
  Naturally, a [Zn/Fe] measurement based on  more than two ions will lead to a better understanding of the depletion.% and also a better understanding of nucleosynthesis effects.  
These [Zn/Fe] constraints  will determine the amount of dust in the CGM, which is itself well correlated with metallicity \citep[Fig.6 of][]{DeCia:2016a}, and our aim is to determine whether the dust/metallicity content of the CGM differs on the minor/major axes of galaxies. 
The information in \citet{DeCia:2016a} were also applied in \citet{Guber:2018a} in using
 \MnII/\CaII{} as an indicator for the dust-to-gas ratio in Damped Lyman Alpha (DLA) systems.
This approach was also followed by \citet{Jones2018}, who extended the prescription to estimate the Ni depletion sequence and assume low ions from different elements to be cospatial as well. 
\citet{Churchill:2015} also conclude from hydroART simulations (at $z\sim0.5$)
of the CGM that low ionization gas likely arises
from small structures with a narrow range of
densities and temperatures, suggesting they can be
modeled as single phase structures. FOGGIE simulations of the CGM at higher redshifts ($2 \lesssim z \lesssim 2.5$)
also find that low ions populate similar regions in velocity phase-space \citep{Peeples:2019}.
The elements used in this study have nearly identical ionization potentials (within 1--2 eV) and
therefore should be strongly associated.
%=======================

%=======================
In our study we do not apply any ionization corrections.
While ionization corrections for an element relative to hydrogen are important at $\log N_{\HI}<19.5$, the corrections for low-ionization elements (e.g. FeII) to others (e.g. ZnII) are relatively small. 
The low-ions we used in this work have similar ionization potential and are, to a large extent, already fully ionized. This is confirmed by studies such as \cite{Peroux:2007} in which they state that for sub-DLAs only a small fraction required ionization corrections above 0.2 dex and none beyond 0.35 dex.
\cite{Meiring:2009} also conclude that the ionization corrections for most elements in DLA systems were found to be < 0.2 dex in most cases. 
We note that \citet{DeCia:2016a} used DLA and  sub-damped Lyman $\alpha$ (sub-DLA) systems
with $N_{\HI}$ > $10^{20}$ cm$^{-2}$ where ionization corrections are negligible and in general below 0.3 dex \citep[e.g.][]{Vladilo:2001a,Dessauges:2003}.
The systems of our sample have  estimated column densities of \HI{} > $10^{19.5}$ cm$^{-2}$ based on the restframe EWs of the \MgII{} absorption \citep{Menard:2009a, Lan:2017}. We made calculations on the required ionization correction for \ZnII{}, \MnII{} and \CrII{} and confirm that the corrections for [Zn/Fe] and certainly [Mn/Fe] and [Cr/Fe] do not exceed 0.2 -- 0.3 dex for
$\log N_{\HI}>18.0$ and $\log N_{\HI}<20.5$ % ($10^{18.0}$ cm$^{-2} \lesssim \HI{} \lesssim  10^{20.5}$ cm$^{-2}$ ) 
and log n$_H  \gtrsim$ -0.5, respectively.

Some of our targets show \MnII{}, \CrII{} and \ZnII{} together. For them, the model fit shows no sign
of systematic discrepancies for \ZnII{} that could be caused by  ionization effects.
Thus, while such ionization effects --if present-- could introduce a slight offset in our column density measurements, the column densities would be shifted by a similar amount and have little impact on the derived slope and hence on [Zn/Fe] (see \Figs{fig:depletion:examples}).
Moreover, it should be noted that these corrections are heavily model dependent. Another source of error is from the mis-association between \HI(v) and X(v) and multi-phase nature of the gas.

Our calculations show that even under strongly varying conditions with regard to gas density and metallicity the required corrections for [Zn/Fe] are significantly lower than the observed range of > 1 dex.

\section{Results}
\label{sec:results}
\subsection{Dust anisotropy}

The primary goal of this study is to investigate the relationship between the 
(known) anisotropic distribution of MgII around galaxies \citep[e.g.,][]{Bouche:2012a, Kacprzak:2012a, Lopez:2018} with the distribution of dust in the CGM.

\begin{figure*}
\centering
\includegraphics[width=0.7\textwidth]{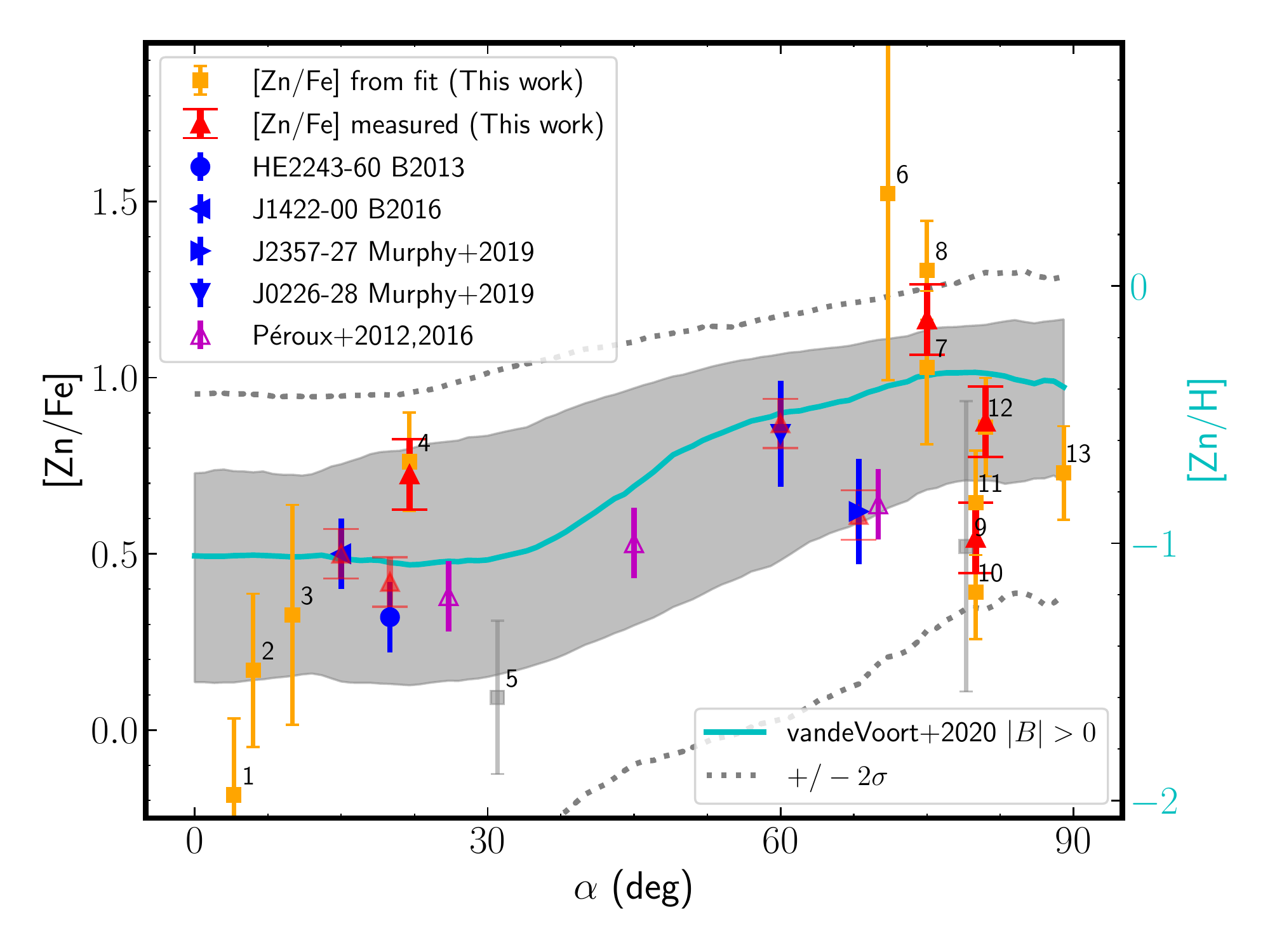}%ZnFefit_tmp.png}
\caption{[Zn/Fe] depletion as a function of azimuthal angle $\alpha$ from our galaxy-quasar pairs from MEGAFLOW using the depletion methodology of \citet{DeCia:2016a} described in section~\ref{sec:methodo}. The labeled IDs and their corresponding systems are listed in Table \ref{table:targets}. For the right axis, the relation for a constant dust-to-gas ratio from \citet{DeCia:2016a} is used ([Zn/H] = 1.37 $\times$ [Zn/Fe] - 1.76).The cyan line represents the model from \citet{Voort:2020a} with non-zero magnetic field with their $1\sigma$ ($2\sigma$) error as shaded (dotted lines), respectively. Individual literature data are shown from \citet{Peroux:2012,Peroux:2016a} (open triangles) and from \citet{Bouche:2013a,Bouche:2016a,Murphy:2019} (blue symbols, see also Table \ref{tab:literature}). The red symbols are direct measurements using Zn and Fe.  The grey data points are possibly affected by saturation in \MgII{}. }
\label{fig:depletion:ZnFit}
\end{figure*}

Figure \ref{fig:depletion:ZnFit} shows the depletion [Zn/Fe] values as a function of the azimuthal angle $\alpha$ for our subset of galaxy-absorber pairs from MEGAFLOW. The individual systems are shown with the solid squares and are labeled by their ID in order of azimuthal angle (see Table \ref{table:targets}).
The red solid triangles show the direct [Zn/Fe] measurement when available.
  One sees that systems located along the minor-axis of galaxies, with $\alpha\goa60$, have depletion levels of  [Zn/Fe] $\approx1.0$, while systems located along the major-axis of galaxies,  with $\alpha\loa30$, have smaller depletions with [Zn/Fe] at 0--0.5, indicative of a smaller dust content.

In Figure \ref{fig:depletion:ZnFit}, we also show literature data when [Zn/Fe] is available and  $\alpha$ well defined,
which are taken from \citet{Peroux:2016a}~\footnote{The azimuthal angles $\alpha$ and column densities come from \citet{Peroux:2016a} and \citet{Peroux:2012}, respectively.} (open triangles),  from \citet{Bouche:2013a, Murphy:2019} (blue triangles) and from \citet{Bouche:2016a} (blue circles).
For \cite{Bouche:2013a} and \citet{Bouche:2016a}, we were able to perform the same analysis as in this paper using the same modeling.

\begin{figure}
\centering
\includegraphics[width=0.45\textwidth]{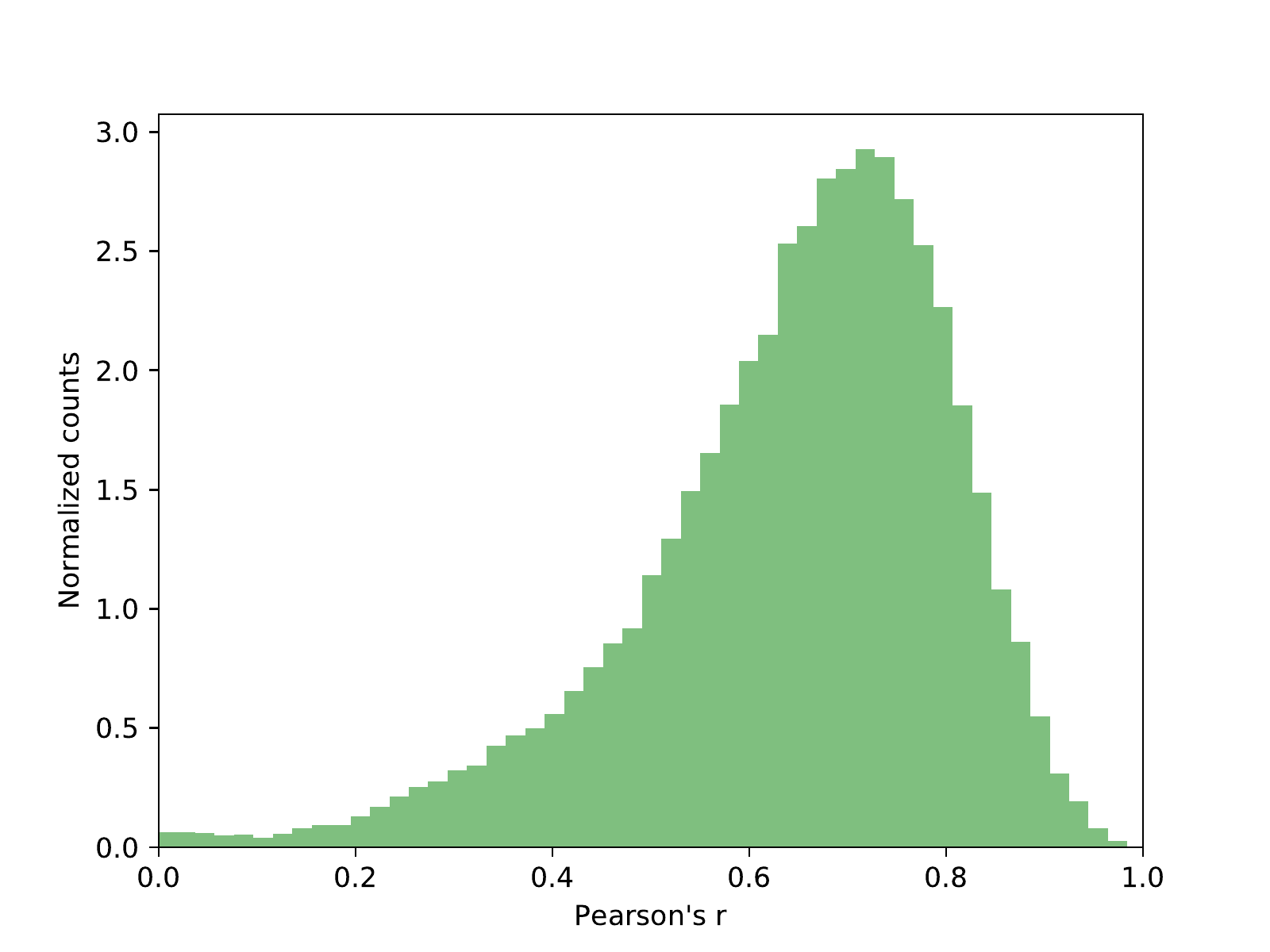}
\caption{The distribution of the correlation coefficient for [Zn/Fe] against azimuthal angle for 50\,000 bootstrap samples of the data plotted in Fig.~\ref{fig:depletion:ZnFit}.}
\label{fig:bootstrap}
\end{figure}

The found correlation with the measured azimuthal angle $\alpha$ for our data points is significant at more than $4\sigma$ from a bootstrap sampling of the Pearson correlation coefficient. Indeed,
Figure \ref{fig:bootstrap} shows the distribution of the Pearson correlation coefficient of [Zn/Fe] over $\alpha$ for 50\,000 bootstrap realizations $r = 0.70 \pm 0.14$.
This relates well to the direct Spearman correlation with the standard deviation in $r$ as $\sigma_r=\frac{1-r^2}{\sqrt{N-1}}$ and $r = 0.65 \pm 0.17$.

To evaluate the robustness of the results and their interpretation we performed several checks for alternative underlying correlations
that could mimic a dependency between dust depletion and azimuthal angle. Firstly, we wanted to make sure that the trend we see is not attributable to the stellar masses of the galaxies in our sample. The two-sided hypothesis test, whose null hypothesis is that these two properties are uncorrelated, yields
a p-value of 0.49 for $M\star$ vs $\alpha$.
Similarly, the [Zn/Fe] values show no correlation with the impact parameter to the galaxy center (p-value for null hypothesis is 0.97). The values for $M\star$ are also listed in Table \ref{table:targets}. For consistency, the given masses are
estimated via the galaxies' dynamics using the S$_{\mathrm{05}}$-$M\star$ relation from \citet{Alcorn:2018} as utilized in \citet{Schroetter:2019a} and therefore differ slightly
from the masses derived via SED-fittig in \paptwo{} 
for the accretion cases.
Our sample covers a wide range of [Zn/Fe] (-0.2 < [Zn/Fe] < 1.5) and relative galaxy orientations ($\alpha$ = 4 - 89 degrees). A potential bias towards enriched gas 
could not mimic the observed trend.
\begin{figure}
\centering
\includegraphics[width=0.5\textwidth]{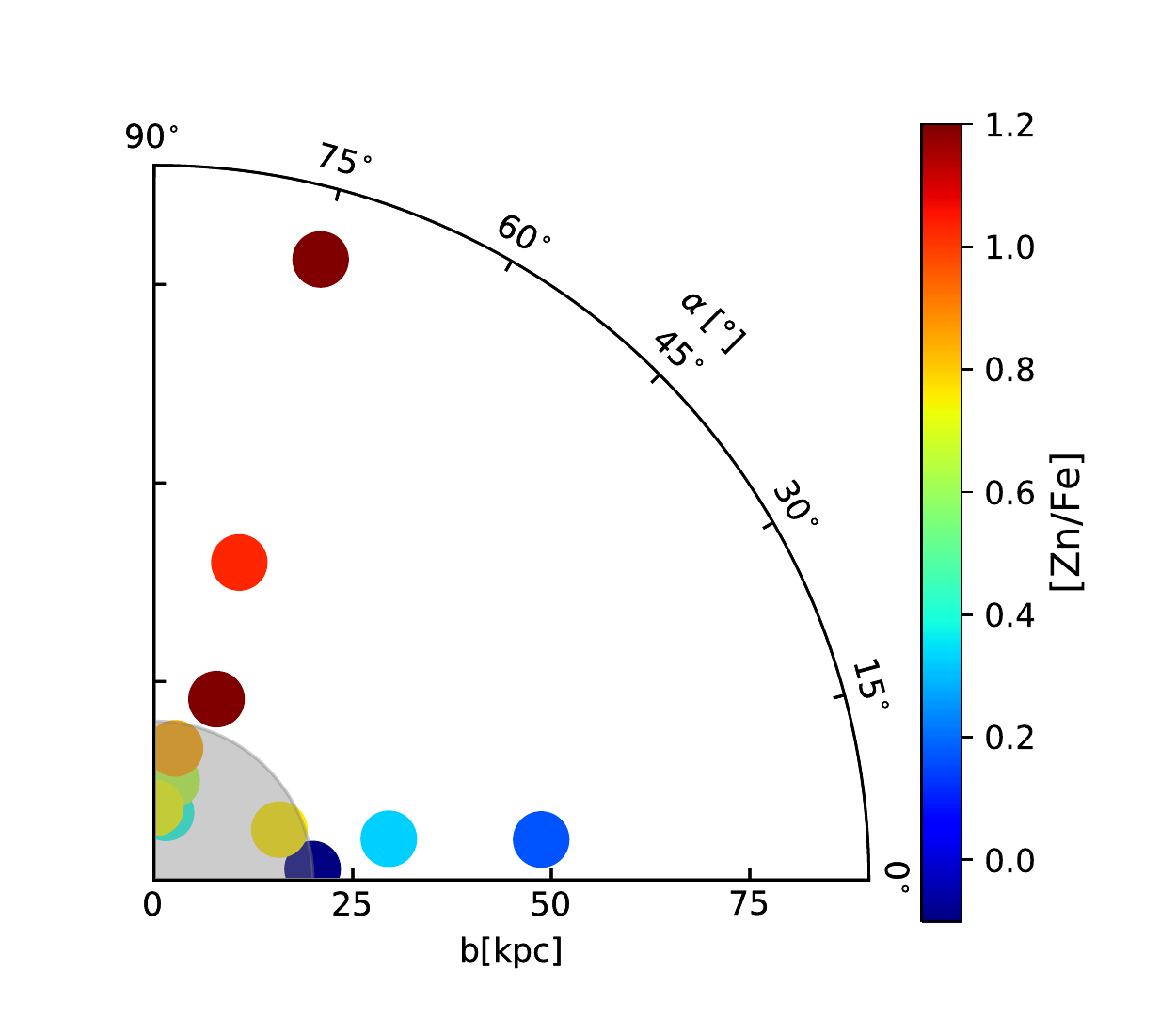}
\caption{The position of the sightlines given as azimuthal angle $\alpha$ at the distance (in kpc) and the obtained [Zn/Fe] color-coded. Blue data points represent lower dust content ([Zn/Fe]<0.4), mostly for lower angles, aligned to the major axis) red represents a higher metallicity ([Zn/Fe]>0.8, mostly along the minor axis at higher azimuthal angles). The grey circle indicates the ISM region up to 20 kpc.}
\label{fig:galaxy}
\end{figure}
\begin{table*}
\caption{Details of the absorber-galaxy pairs. The uncertainties of the fitted [Zn/Fe] (given in parentheses referring to the corresponding last digits of the quoted result) reflect the plain fitting error. For the model fit we assumed a lower limit of 0.1 dex for the uncertainty in column density. The pair ID is also shown in Fig.~\ref{fig:depletion:ZnFit}. The rounded impact parameter b, azimuthal angle  $\alpha$ and $M\star$ are from \paptwo{} and \papthree. }
\begin{tabular}{l|l|l|c|c|c|c|c|c|}
  \hline
  Pair ID & Quasar ID & redshift & b [kpc]& $\alpha$ [deg]& $W^{\lambda2796}_{\rm r}$ [\r{A}] & [Zn/Fe]$_{\rm obs}$ & $M\star$ & [Zn/Fe] model \\
  \hline
  1 & J0103+1332 & 0.788 & 20 & 4 & 1.25 & -- & 9.1 & -0.18(22) \\
  2 & J2152+0625 & 1.053 & 49 & 6 & 0.54 & -- & 10.1 & 0.17(22) \\
  3 & J1358+1145 & 1.418 & 30 & 10 & 2.61 & -- & 8.9 & 0.33(31) \\
  4 & J1236+0725 & 0.912 & 17 & 22 & 2.24 & 0.73 & 10.5 & 0.76(14) \\
  5 & J1107+1021 & 0.745 & 43 & 31 & 2.45 & -- & 8.4 & 0.09(22) \\
  6 & J0015-0751 & 0.507 & 24 & 71 & 1.59 & -- & 10.7 & 1.52(53) \\
  7 & J0937+0656 & 0.933 & 41 & 75 & 1.52 & -- & 9.7 & 1.03(22) \\
  8 & J1107+1021 & 1.015 & 81 & 75 & 1.93 & 1.16 & 11.1 & 1.30(14) \\
  9 & J0800+1849 & 0.843 & 21 & 79 & 0.96 & -- & 9.8 & 0.52(41) \\
  10 & J1039+0714 & 1.359 & 9 & 80 & 2.49 & -- & 9.0 & 0.39(13) \\
  11 & J1358+1145 & 0.810 & 13 & 80 & 1.79 & 0.55 & 9.3 & 0.64(15) \\
  12 & J1358+1145 & 0.704 & 17 & 81 & 2.45 & 0.88 & 10.2 & 0.86(14) \\
  13 & J0103+1332 & 1.048 &  9 & 89 & 2.99 & -- & 9.8 & 0.73(13) \\
  \hline
 \end{tabular}

\label{table:targets}
\end{table*}
\subsection{Implications for CGM Metallicity}
\label{sec:metallicity}

The implication of our result is that the CGM metallicity ought to also be anisotropic, provided that
there is a relation between dust depletion and metallicity.
At face value, there seems to be such a dust-metallicity ([Zn/Fe]--[Zn/H]) relation, albeit with significant intrinsic scatter
\citep{Meiring:2009,DeCia:2016a}.

In this context, we also show in Figure~\ref{fig:depletion:ZnFit} (cyan line; the right-hand $y$-axis) the recent  prediction of the azimuthal dependence of metallicity from a simulation of a MilkyWay-like halo that included magnetic fields from \cite{Voort:2020a}~\footnote{The model was arbitrarily shifted by -0.5 dex in X/H
 given the uncertain yields in the simulations (F. van de Voort, private communication).}.
The right-hand $y$ axis is computed from the   [Zn/Fe]--[Zn/H] relation of \citet{DeCia:2016a}.

Provided that there is a dust-metallicity relation,
our results are in apparent contradiction with the recent analysis of \citet{Pointon2019}  who studied
the azimuthal distribution of the CGM metallicity for the first time, using a sample of 47 galaxies with  \HI\ absorption  ranging from $10^{13.8}$ - $10^{19.9}$ cm$^{-2}$ at a considerably lower median redshift of  $z\sim0.3$ detected in 29 quasar fields. 
Using photo-ionization models, they find no evidence for a relation between their single phase metallicities and $\alpha$ on 29 systems.
 However, we note that the impact parameter of the
galaxies in their sample are rather large (median impact parameter $\sim$ 93 kpc) with only 10 galaxies having impact parameter of < 50 kpc.
\section{Conclusion \& Discussion} 
\label{sec:conclusions}
In order to study the dust-depletion pattern of the CGM gas around galaxies from our MEGAFLOW survey,
we selected all the absorber-galaxy pairs where column densities of at least two elements such as Zn, Fe, Mn, etc. could be measured in our UVES spectra.
From this sub-sample, we find that the dust-depletion ([Zn/Fe]) does depend strongly on the azimuthal angle of the quasar's apparent location with respect to the galaxy's major-axis (see Fig~\ref{fig:depletion:ZnFit}).
Fig.~\ref{fig:galaxy} shows the same data as Fig~\ref{fig:depletion:ZnFit} but includes the information of the impact parameter and thus
corresponds to a 2D projection of a quadrant of the average galaxy we see in CGM absorption (without any information on inclination).
Given that [Zn/Fe] is also a proxy for metallicity \citep[as discussed in][]{DeCia:2016a} where [Zn/Fe]=0 (1) corresponds to sightlines with [M/H] $\approx$-1.75 (-0.25), respectively, one can conclude that the bi-modal distribution of \MgII{} absorbers around galaxies is indeed related to metal-poor gas along the major-axis and metal-rich gas along the galay's minor-axis, supporting the accretion/wind dichotomy.

There are several recent efforts to address the azimuthal dependence in contemporary simulations.
\citet{Kannan:2020} find on the basis of high resolution hydrodynamical simulations of Milky-Way like and Large Magellanic Cloud like  galaxies that dust is efficiently entrained and expelled out of the disk by supernova driven outflows, offering a path to explain the transport of dust from galaxies into the circum-galactic medium.
\citet{Hafen:2019} study the CGM based on the cosmological hydrodynamic FIRE-2 simulations 
and find that the metallicity of IGM accretion is systematically lower than
the metallicity of winds (typically by >1 dex). While they identify the accretion preferably in the galactic plane at low azimuthal angles, they find no distinct bimodal distribution as the simulated winds are distributed close to spherically in the median. They note, however, that a subset of the simulated haloes have enhanced warm/hot wind mass along the galaxy minor axis, which may reflect the preferential expansion of warm/hot wind normal to the disc plane.

\cite{Hopkins:2020} demonstrate the importance of the inclusion of cosmic rays in simulations for
galaxy outflows. They find that the cosmic ray driven outflows extends outwards biconically
with a widening outflow angle at increasing distance.
Since in their simulations the density in outflowing gas is relatively low,
it is not obvious if the clear bipolar outflow structure they find translates directly
to a clear observable trend of absorber equivalent width as a function of polar angle. This influence of cosmic rays on the structure of gas flows in the CGM is also analysed more general in \citet{Buck:2020}.

\cite{Voort:2020a} study the effect of magnetic fields on the CGM properties (such as metallicity) of a Milky Way-like galaxy and find that magnetic fields have a significant impact on the physical properties of the CGM (resolved with spatial refinement to 1 kpc or better). In particular, magnetic fields enhance the azimuthal dependence of the CGM metallicity (see also Fig. \ref{fig:depletion:ZnFit}) in their simulations. The impact of magnetic fields in simulations is also shown in \citet{Sparre:2020}.
Very recently, \citealt{Peroux:2020a} study the physical properties of the CGM in cosmological hydrodynamical simulations and
find a clear angular dependence of the CGM metallicity. The predicted correlation of metallicity with azimuthal angle was
found to be rather robust within a broad mass range of 8.5 < $M\star$ < 10.5 and for all redshifts z < 1.

Our study finds the first evidence for an anisotropic dust distribution in the CGM, and provided there is a dust-metallicity relation, this observed dust-anisotropy supports the results of \citet{Voort:2020a} and the interpretation of dust
enriched outflows along the minor axis and gas with lower metallicity along the
major axis.

\section*{Acknowledgements}
We thank the referee for detailed comments.
This work has been carried out thanks to the support of the ANR 3DGasFlows (ANR-17-CE31-0017), and the OCEVU Labex (ANR-11-LABX-0060).
SM acknowledges support from the Humboldt Foundation, Germany.  We thank M. T. Murphy for providing some of the data in a tabulated format.
This  work  made  use  of  the  following  open  source
software:
\gpk{} \citep{Bouche:2015a},
\textsc{ZAP} \citep{SotoK:2016b},
\textsc{MPDAF} \citep{Piqueras:2017a},
\textsc{matplotlib} \citep{Hunter:2007a},
\textsc{NumPy} \citep{vandderWalt:2011a, Oliphant:2007},
\textsc{Astropy} \citep{Robitaille:2013a},
\textsc{especia} \citep{Quast:2016}.
\section*{data availability}
The data underlying this article are available in the ESO archive 
(http://archive.eso.org).
The reduced data will be shared on reasonable request to the 
corresponding author.
\bibliographystyle{aa}
\bibliography{main}
\appendix
\section{Appendix}
The column densities for the measured ions used in this paper are listed in Table \ref{table:columndensities}.

Table \ref{tab:literature} shows the azimuthal angles and measured
dust depletion [Zn/Fe] for the additional data points in Fig.~\ref{fig:depletion:ZnFit} taken from literature.

Figure \ref{fig:depletion:examples} shows all observed and used transitions with fit to the model described in \citet{DeCia:2016a}.

\begin{table*}
\caption{Column densities or upper limits for the elements considered for this study.  Same order as Table \ref{table:targets}. Marked as `--' when either not covered by the data or too saturated to derive a reasonable column density. For the model fit we assumed a lower limit of 0.1 dex for the uncertainty in column density.}
\begin{tabular}{l|l|l|c|c|c|c|c|c|c|c|}
\hline

  Pair ID & Quasar ID & redshift & \MnII     & \CrII    & \ZnII      & \FeII & \MgII \\
            &              &       & \multicolumn{5}{c}{log $N$ (cm$^{-2}$)} \\
  \hline
      
    1 &   J0103+1332 & 0.788 & < 11.72   & < 12.32  & < 11.77    & 13.75(2) &  13.85(10) \\
    2 &   J2152+0625 & 1.053 & < 11.64   & < 12.69  & < 12.03    & 13.32(3) &  13.65(10)\\
    3 &   J1358+1145 & 1.418 & 12.82(14) & 13.39(9) & 12.65(38)  &--        &--  \\
    4 &   J1236+0725 & 0.912 & 12.7(3)   & < 12.19  & 12.7(1) &  14.82(2) &--  \\
    5 &   J1107+1021 & 0.745 &--         & < 12.15  & < 11.54    & 13.96(1) &  14.24(10) \\
    6 &   J0015-0751 & 0.507 & 12.87(13) & --       &--          & 14.56(2) &--  \\
    7 &   J0937+0656 & 0.933 & 12.34(2)   & --  & < 11.51    & 13.34(3) &  14.14(10)\\
    8 &   J1107+1021 & 1.015 & 12.12(5)  & < 11.98  & 12.66(2)   & 14.34(2) &--  \\
    9 &   J0800+1849 & 0.843 & 12.14(2)  & < 12.36  & < 11.38    & 14.33(1) &  15.24(47)\\
    10 &  J1039+0714 & 1.359 & 13.46(1)  & 13.83(6) & 13.1(1)   &--        &--  \\
    11 &  J1358+1145 & 0.810 & 12.51(4)  & 13.30(24 & 12.6(1)   & 14.9(14) &--  \\
    12 &  J1358+1145 & 0.704 & 13.33(3)  & < 12.18  & 13.18(3)   & 15.15(3) &--  \\
    13 &  J0103+1332 & 1.048 & 13.51(3)  & 13.90(5) & 13.48(1)   &--        &--  \\
 \hline
 \end{tabular}

\label{table:columndensities}
\end{table*}
\begin{table*}
\caption{Literature data (\citet{Bouche:2013a,Bouche:2016a,Murphy:2019}) for azimuthal angles, impact parameters and observed [Zn/Fe]$_{\rm obs}$ as well as [Zn/Fe] derived from the depletion model in \citet{DeCia:2016a}. }
\begin{tabular}{l|l|c|c|c|c|}
\hline
Quasar ID & redshift & b [kpc]& $\alpha$ [deg] & [Zn/Fe]$_{\rm obs}$ &  [Zn/Fe] model\\
\hline
J1422-0001 & 1.083 & 12 & 15 & 0.5 & 0.5(1)\\
HE2242-60  & 2.32  & 26 & 20 & 0.42 & 0.32(10)\\
J2357-2736 & 0.815 & 7  & 68 & 0.61 & 0.62(10)\\
J0226-2857 & 1.022 & 2  & 60 & 0.87 & 0.85(10)\\
\hline
 \end{tabular}

\label{tab:literature}
\end{table*}
%\label{fig:depletion:examples}
%
\begin{figure*}
\begin{center}
\includegraphics[width=0.3\textwidth]{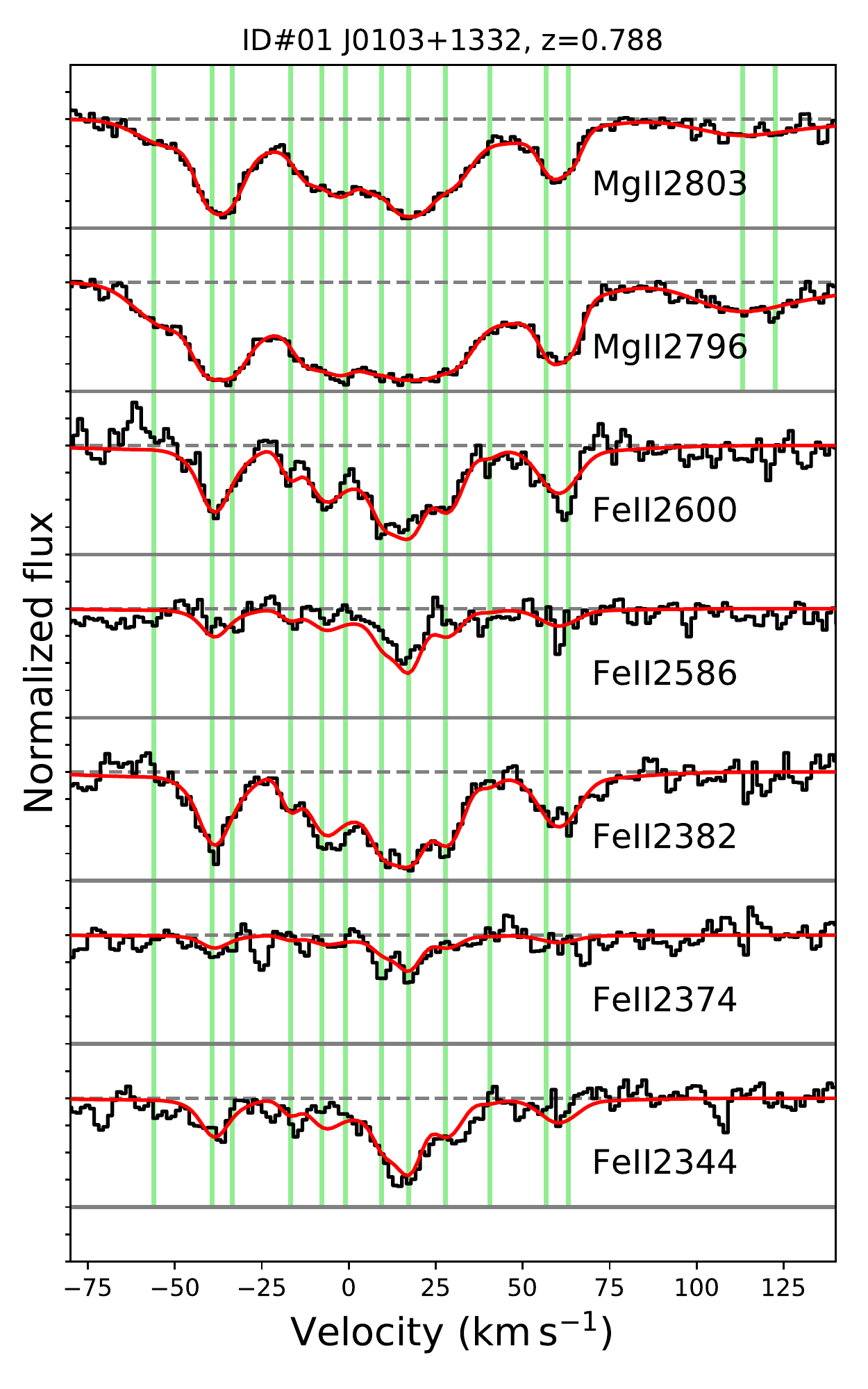}
\includegraphics[width=0.3\textwidth]{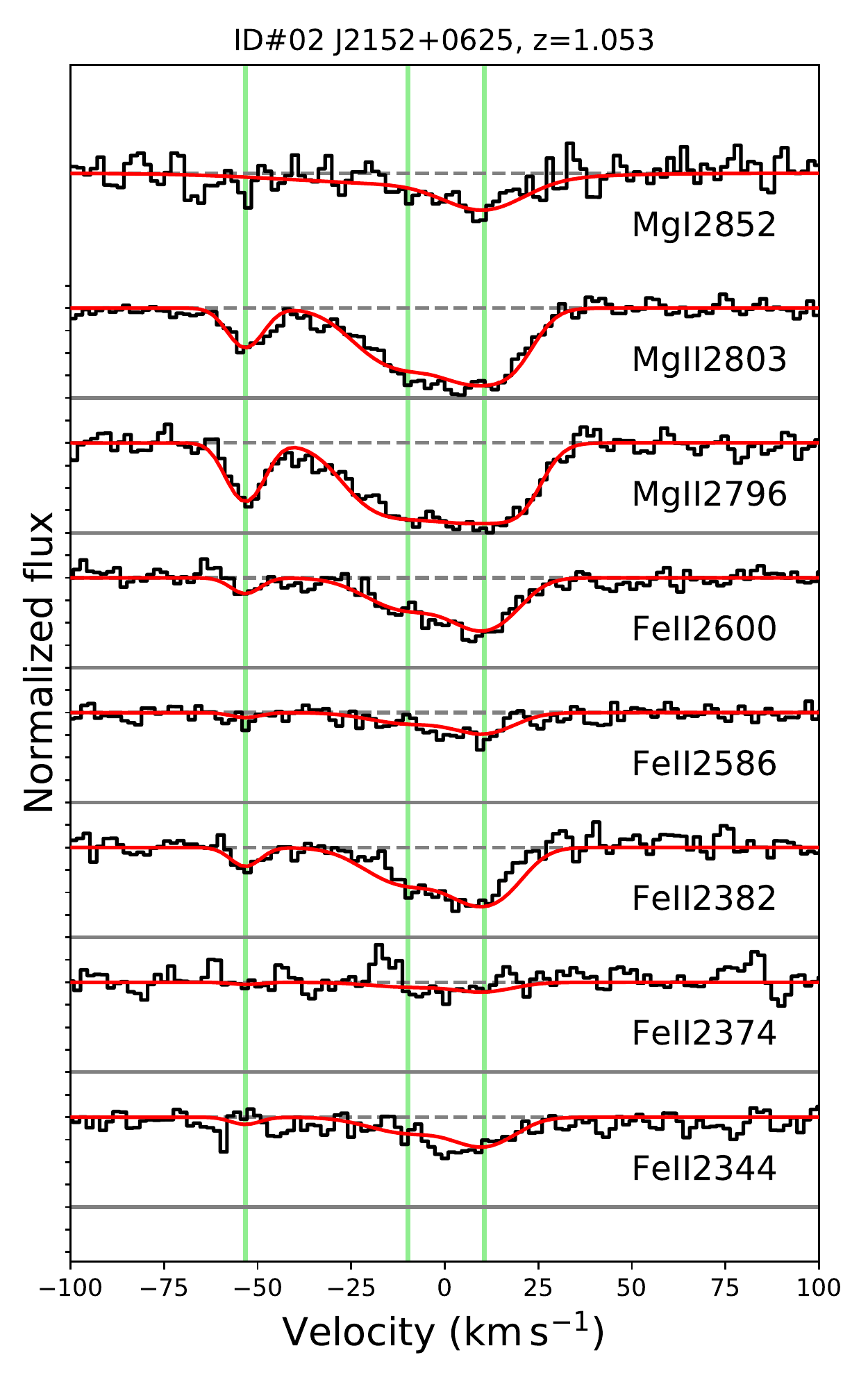}
\includegraphics[width=0.3\textwidth]{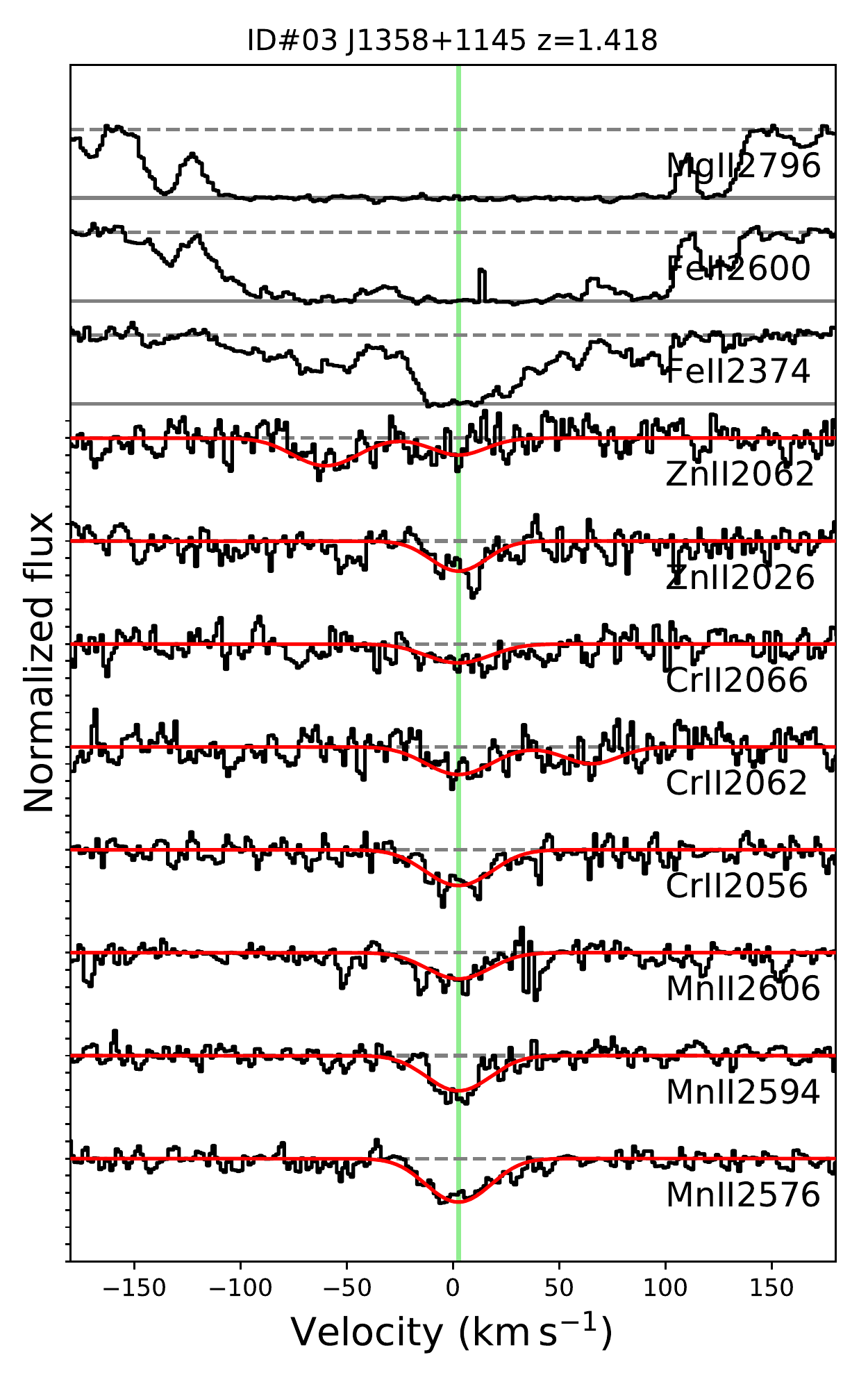}\\
\includegraphics[width=0.3\textwidth]{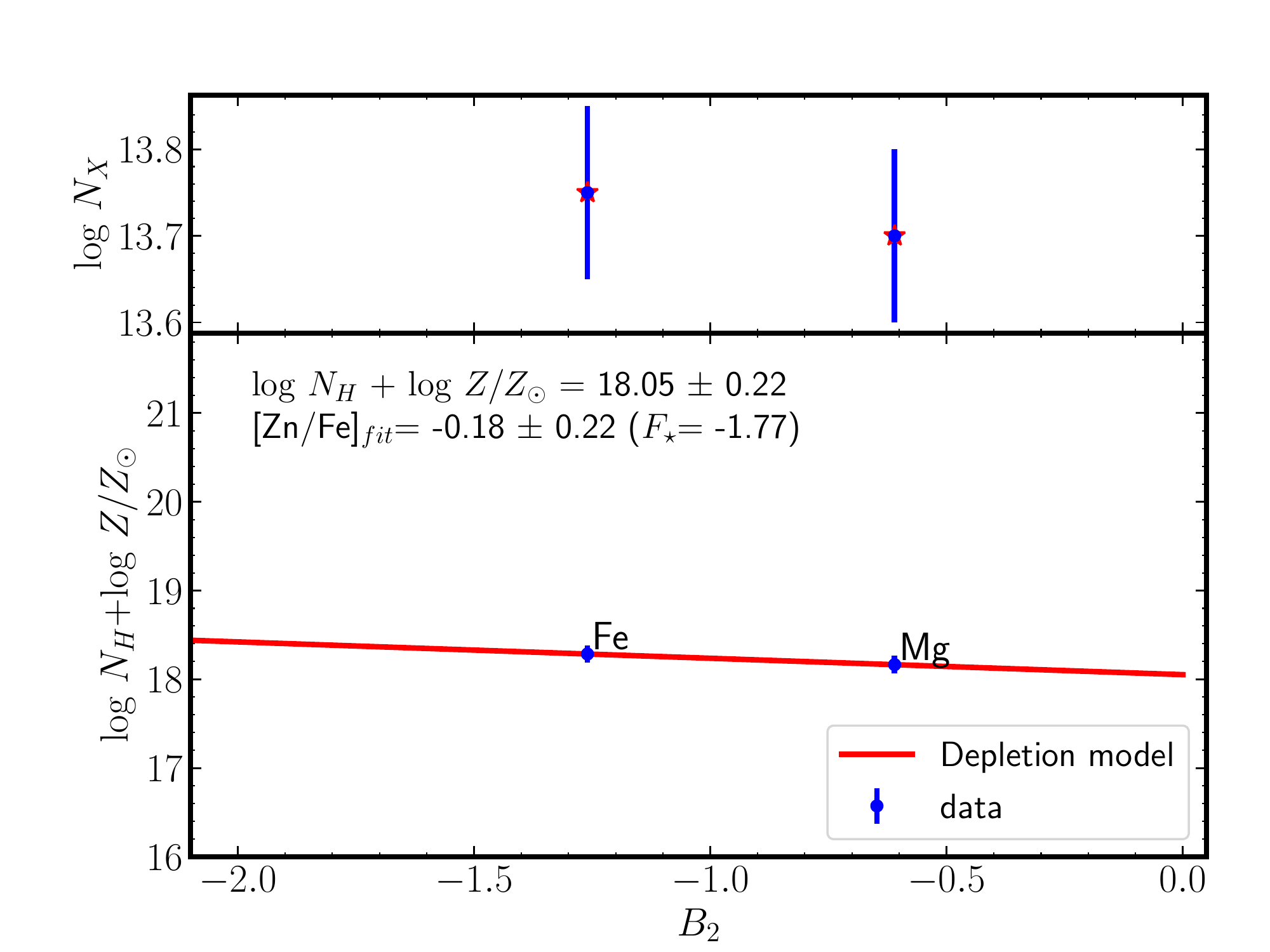}
\includegraphics[width=0.3\textwidth]{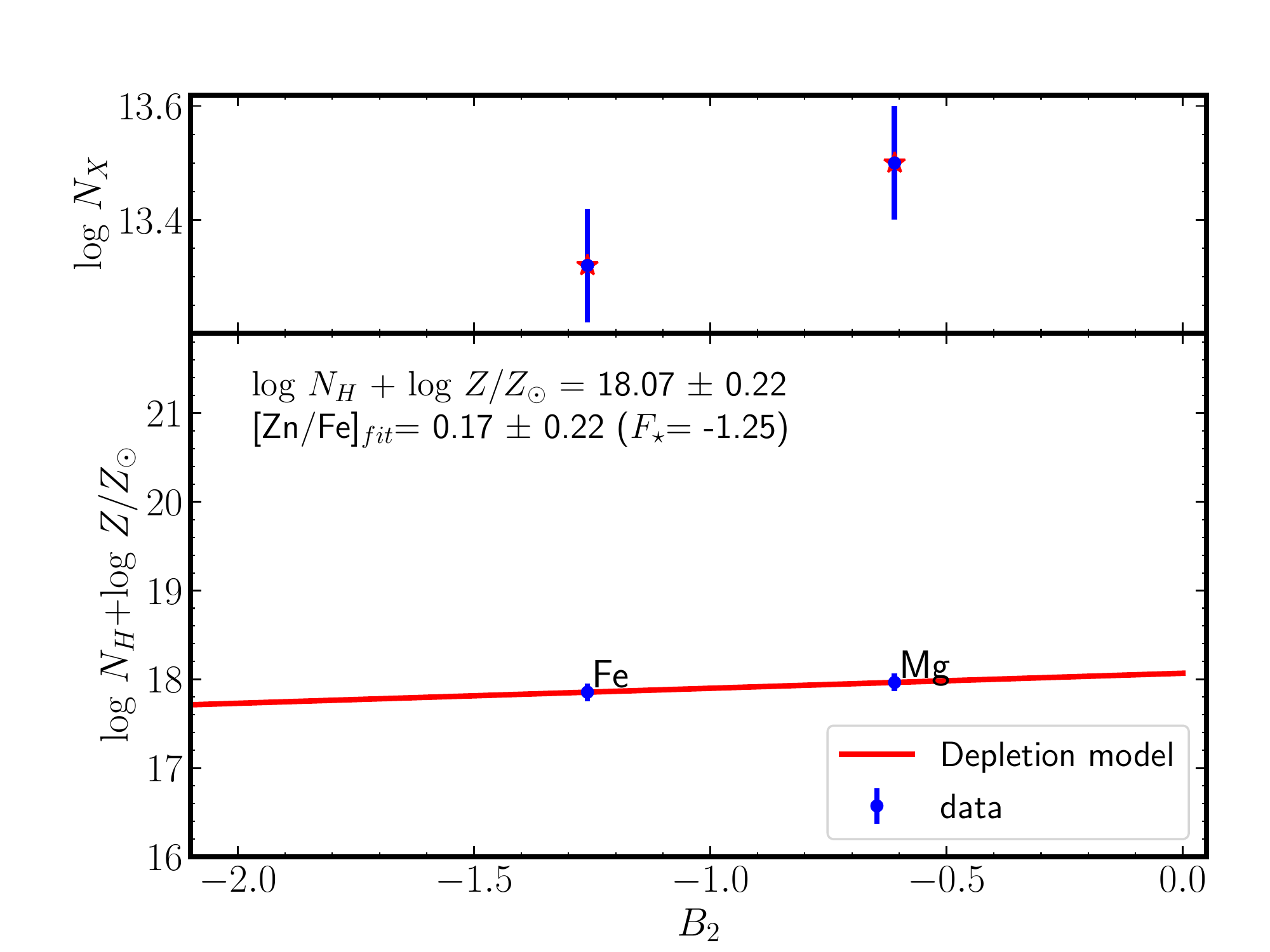}
\includegraphics[width=0.3\textwidth]{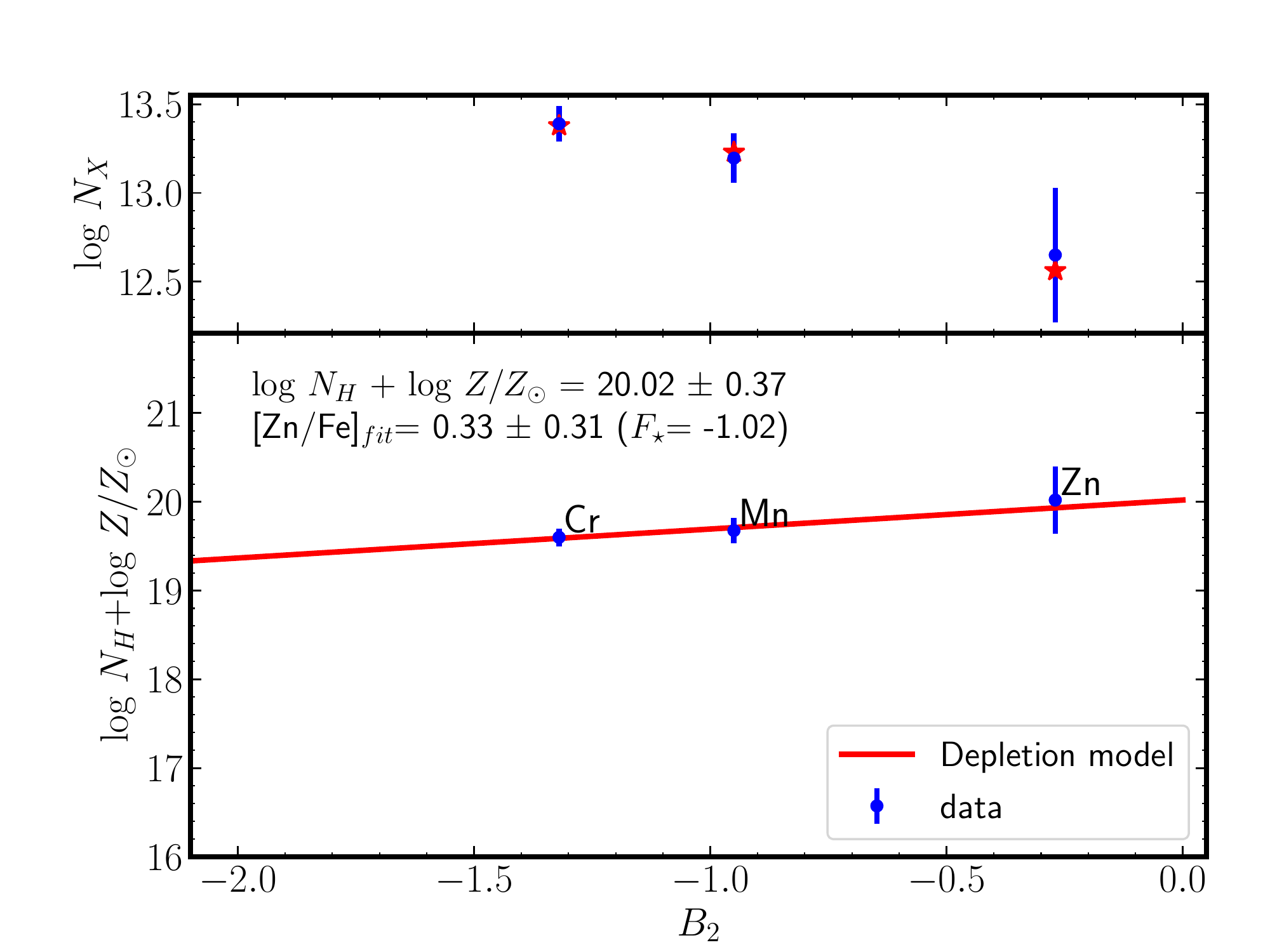}
\caption{Absorption profiles of the measured ions with their corresponding fit ({\it red}). The velocity components
are shown in {\it green}. The dashed and solid {\it grey} lines represent the continuum and zero-level. For very weak
absorption the fluxes are scaled by factor two for this plot. While models are shown for all elements, we derived logN for the weak features from the equivalent width. Below are the fits of the dust depletion scheme of \citet{DeCia:2016a} for the measured ions. Each element (Mn, Zn, Cr, Fe) depletes on dust grains with a different depletion propensity  $B_2$ such that the global depletion level can be solved, and is given by the slope of the red line. The  zero-point at zero depletion ($B_2=0$) gives $\log N_{\mathrm{H}}+\log Z$. The upper panels show the measured column densities ({\it blue}) along with the model predictions ({\it red}).
The global depletion can be parameterized as [Zn/Fe] given that Zn is the least depleted element following \citet{DeCia:2016a} or as $F_\star$ as in \citet{Jenkins:2009a}}
\end{center}
\label{fig:depletion:examples}
\end{figure*}
\begin{figure*}
\begin{center}
\includegraphics[width=0.3\textwidth]{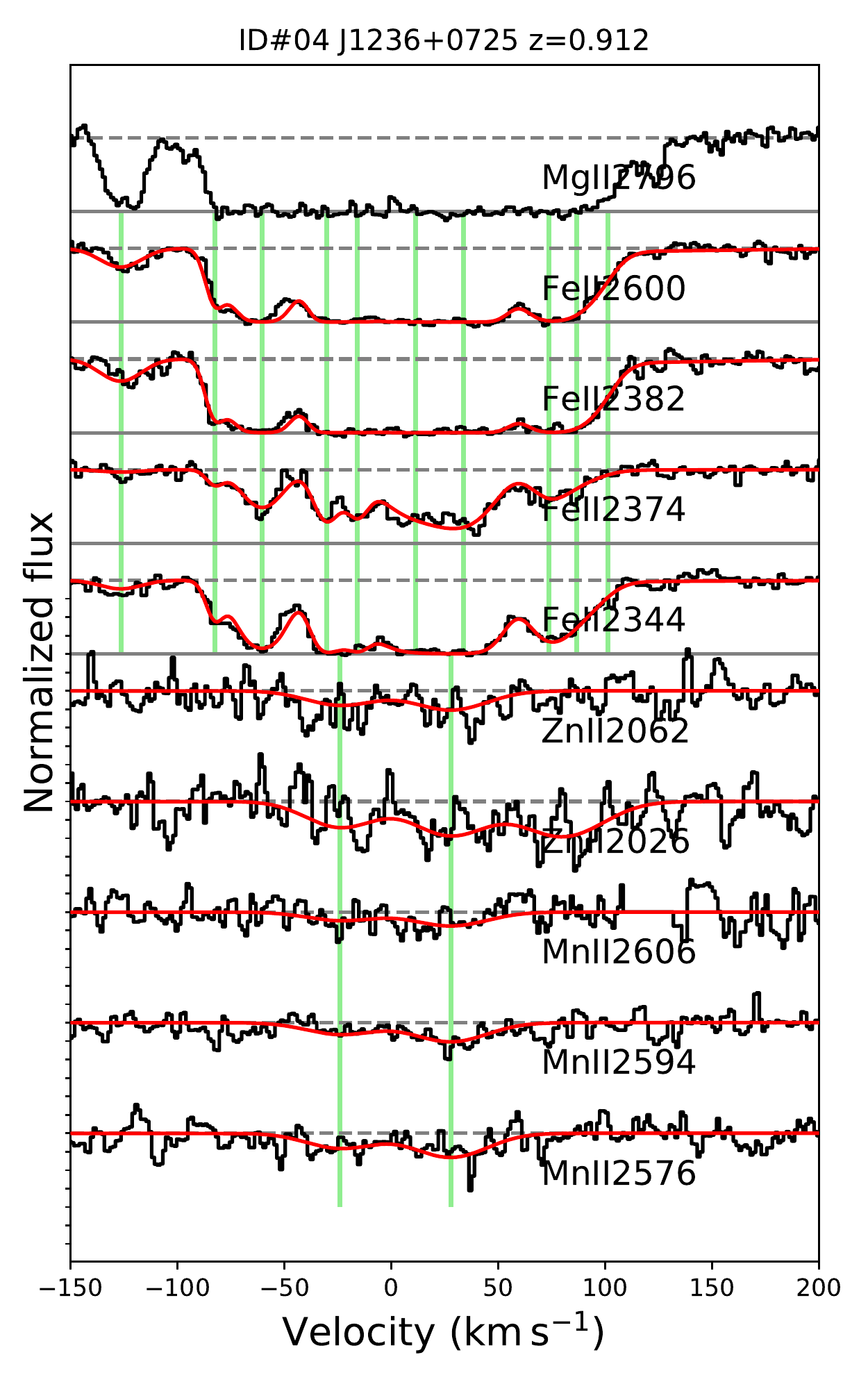}
\includegraphics[width=0.3\textwidth]{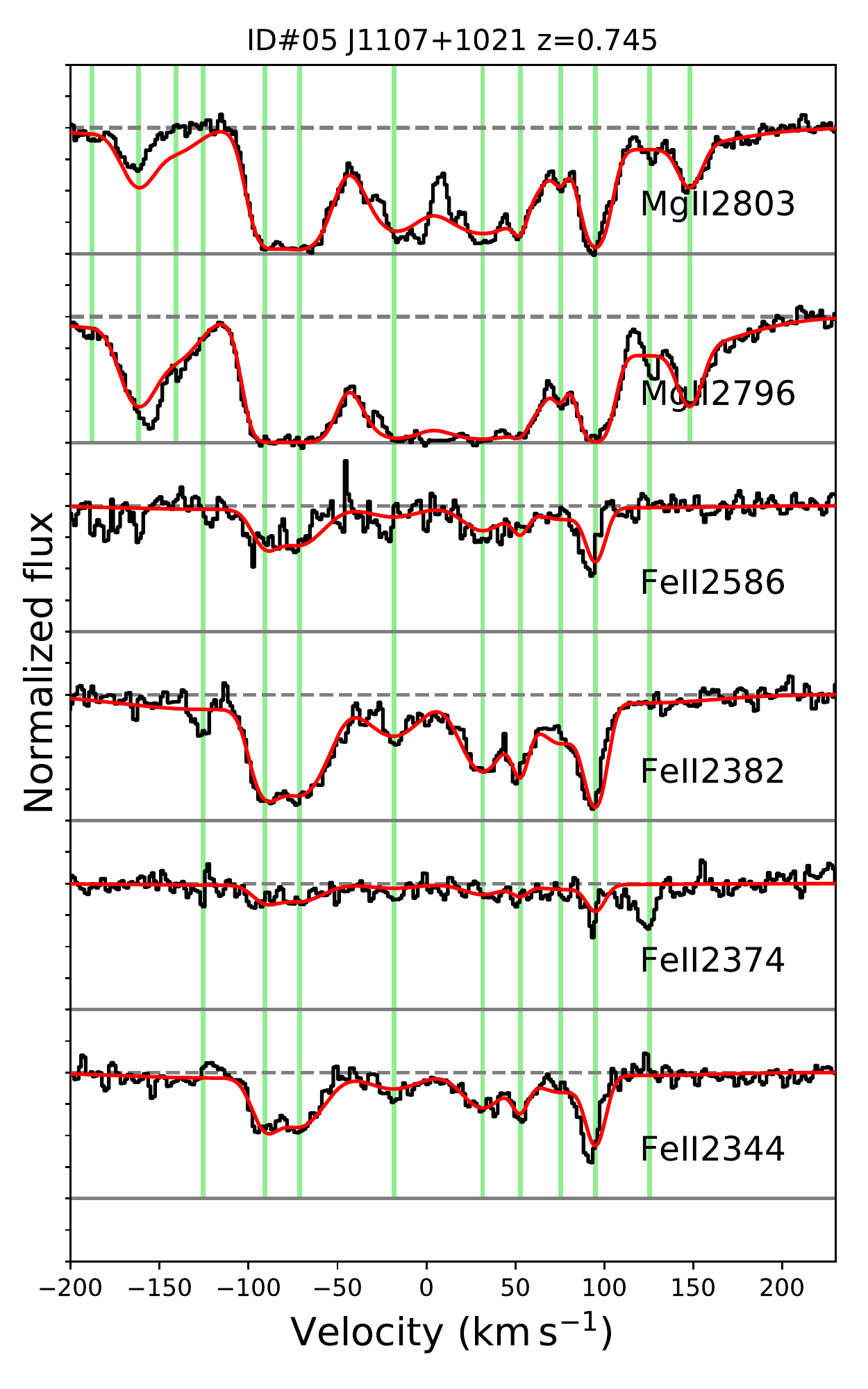}
\includegraphics[width=0.3\textwidth]{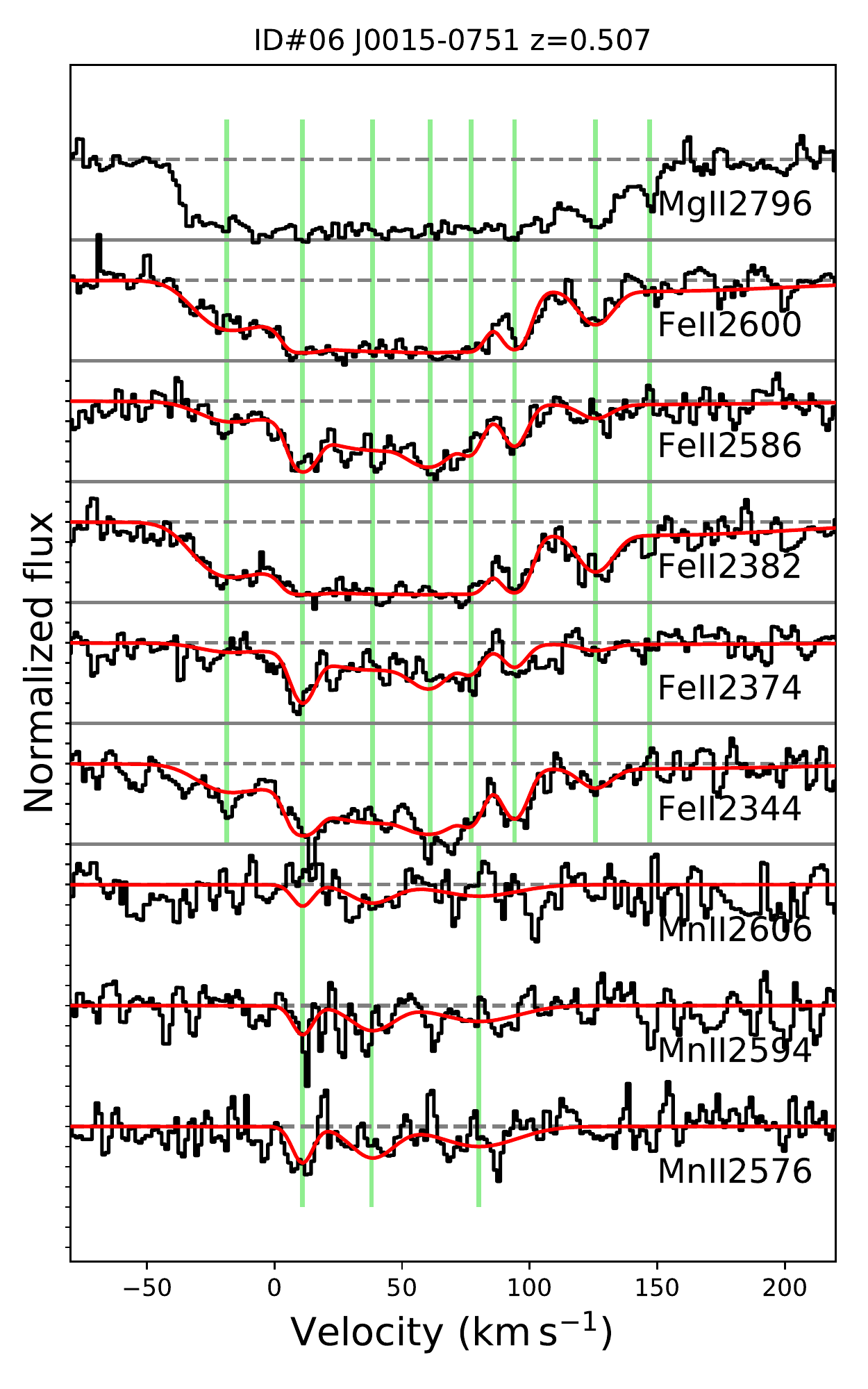}\\
\includegraphics[width=0.3\textwidth]{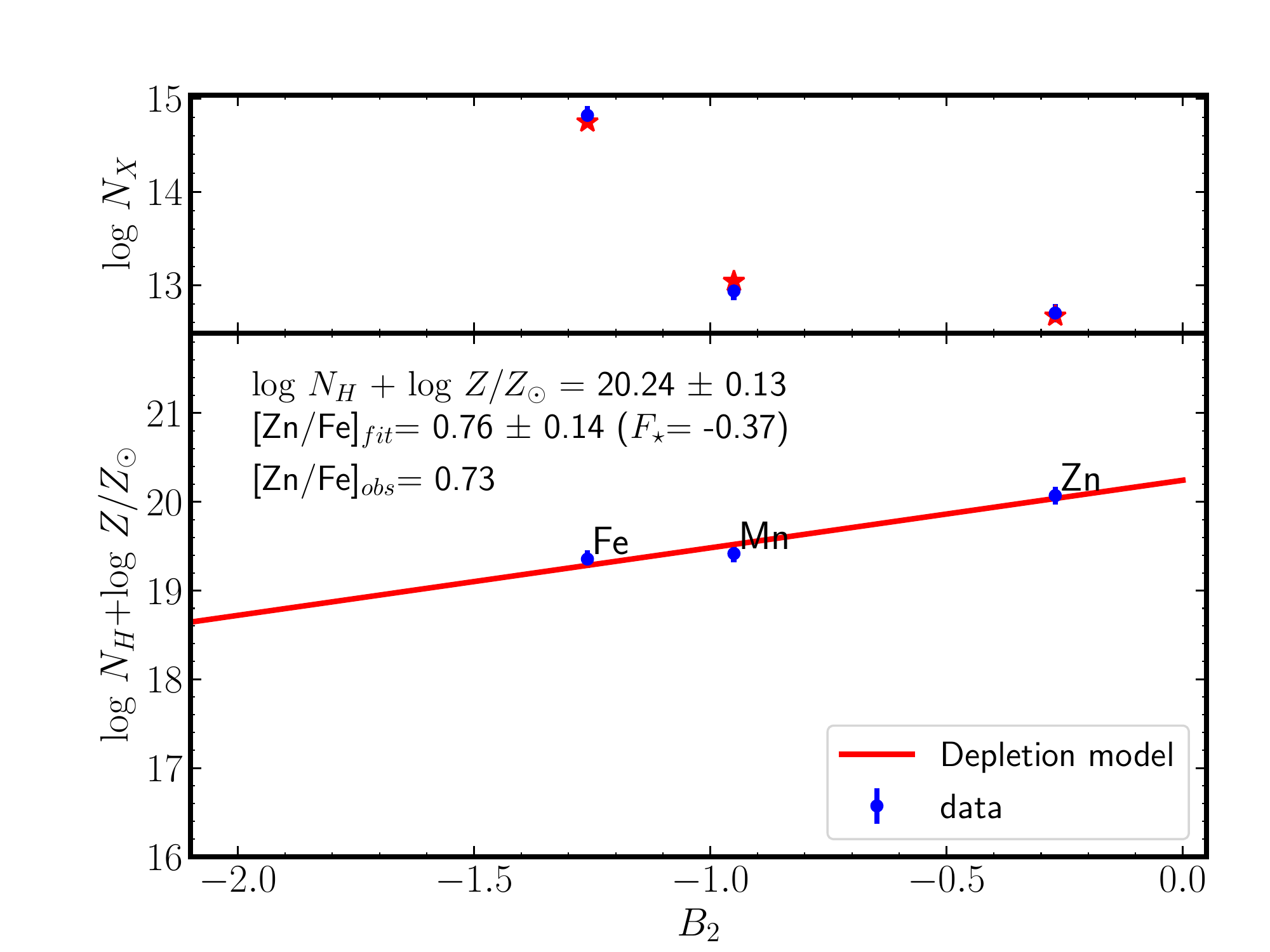}
\includegraphics[width=0.3\textwidth]{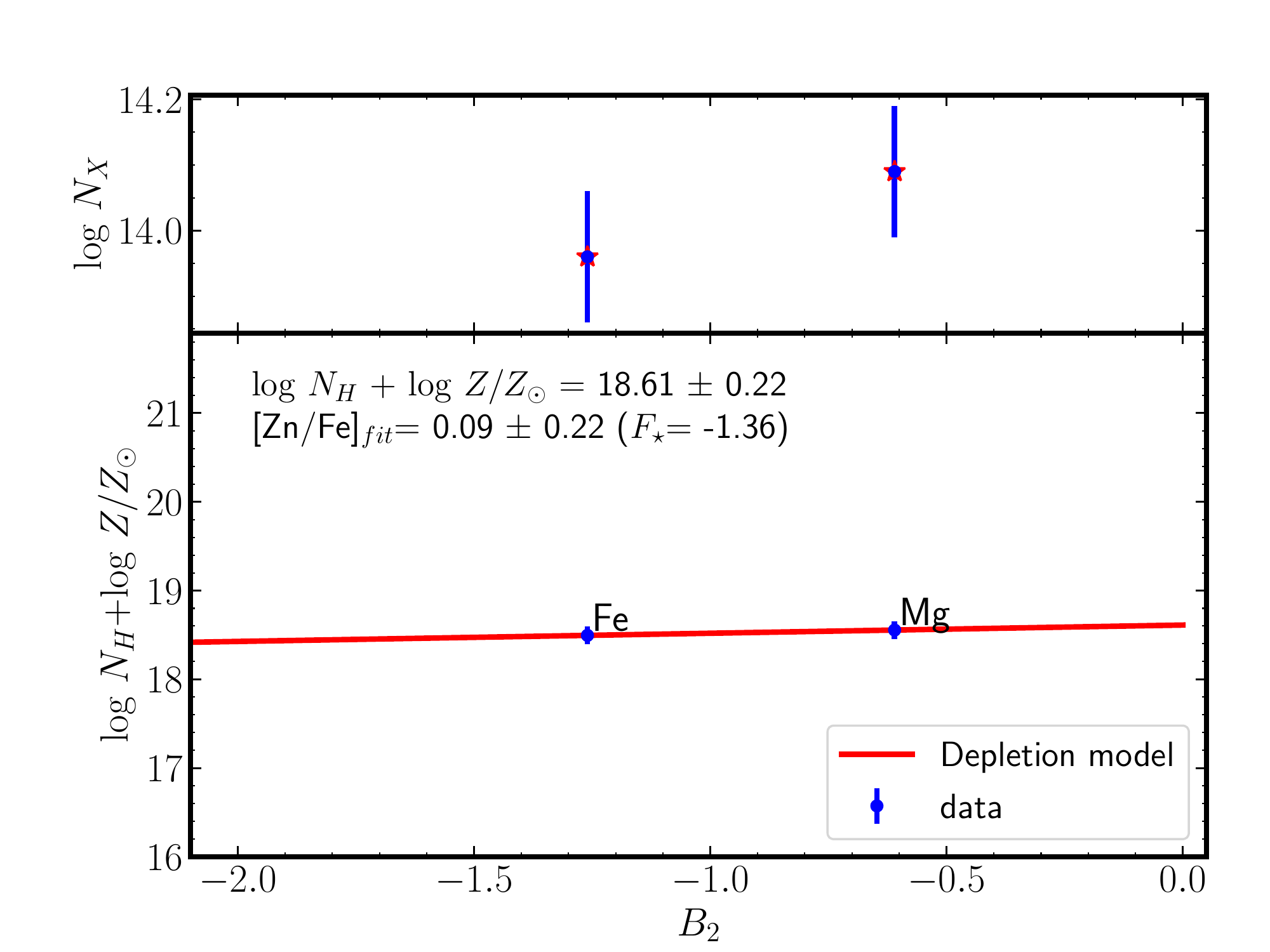}
\includegraphics[width=0.3\textwidth]{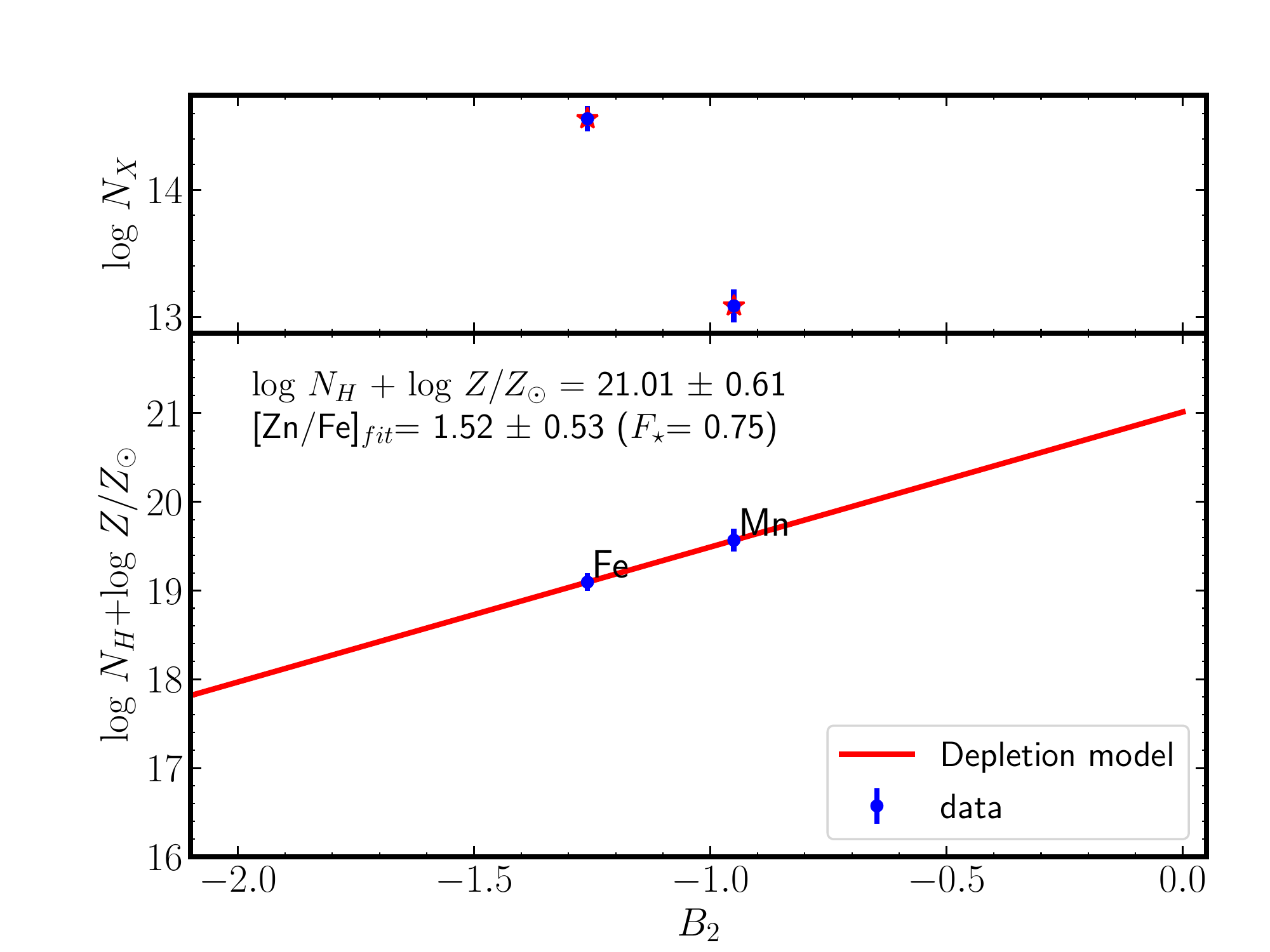}
\end{center}
\contcaption{Continued from Figure \ref{fig:depletion:examples}.}
\label{fig:example_cont}
\end{figure*}
\begin{figure*}
\begin{center}
\includegraphics[width=0.3\textwidth]{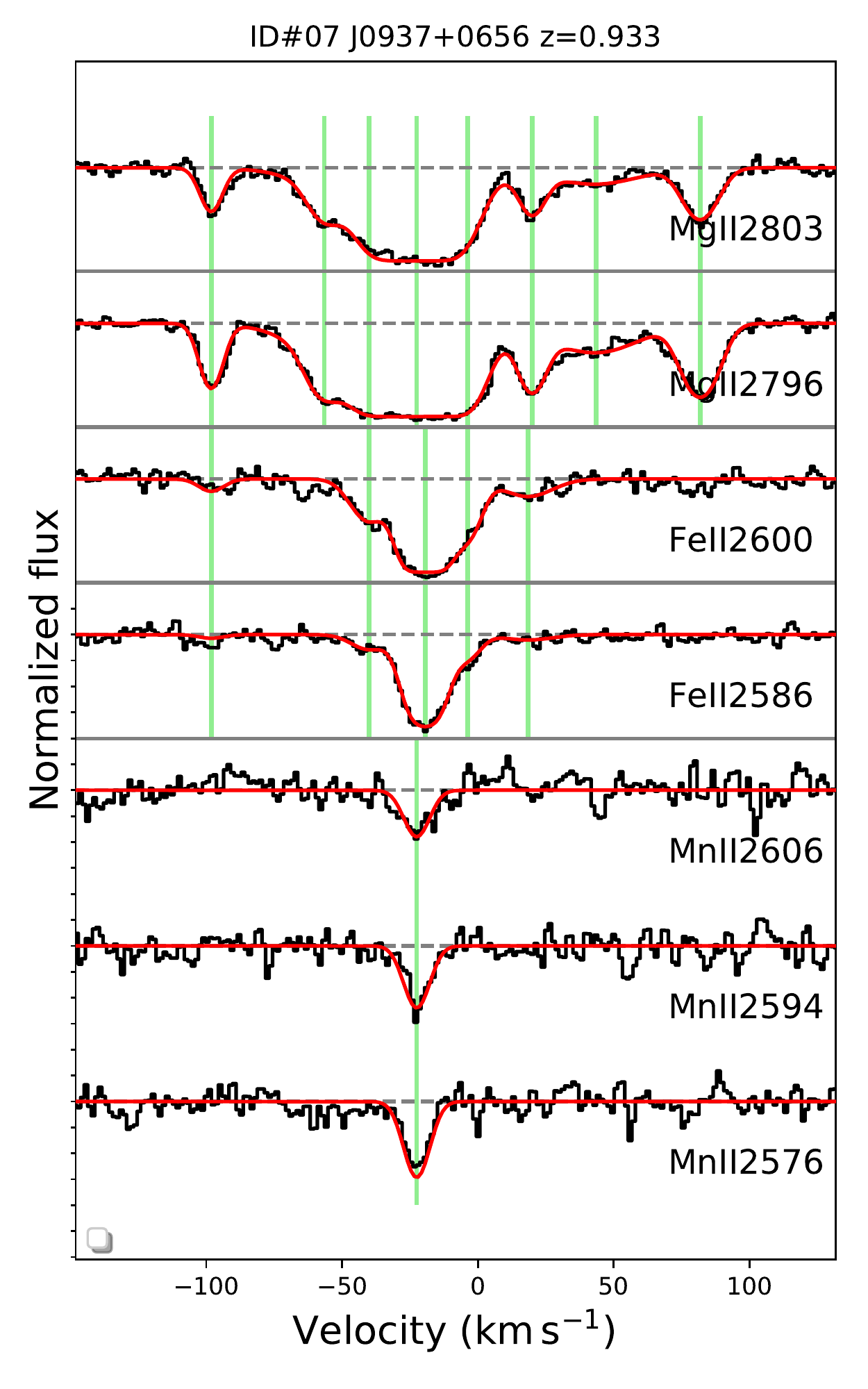}
\includegraphics[width=0.3\textwidth]{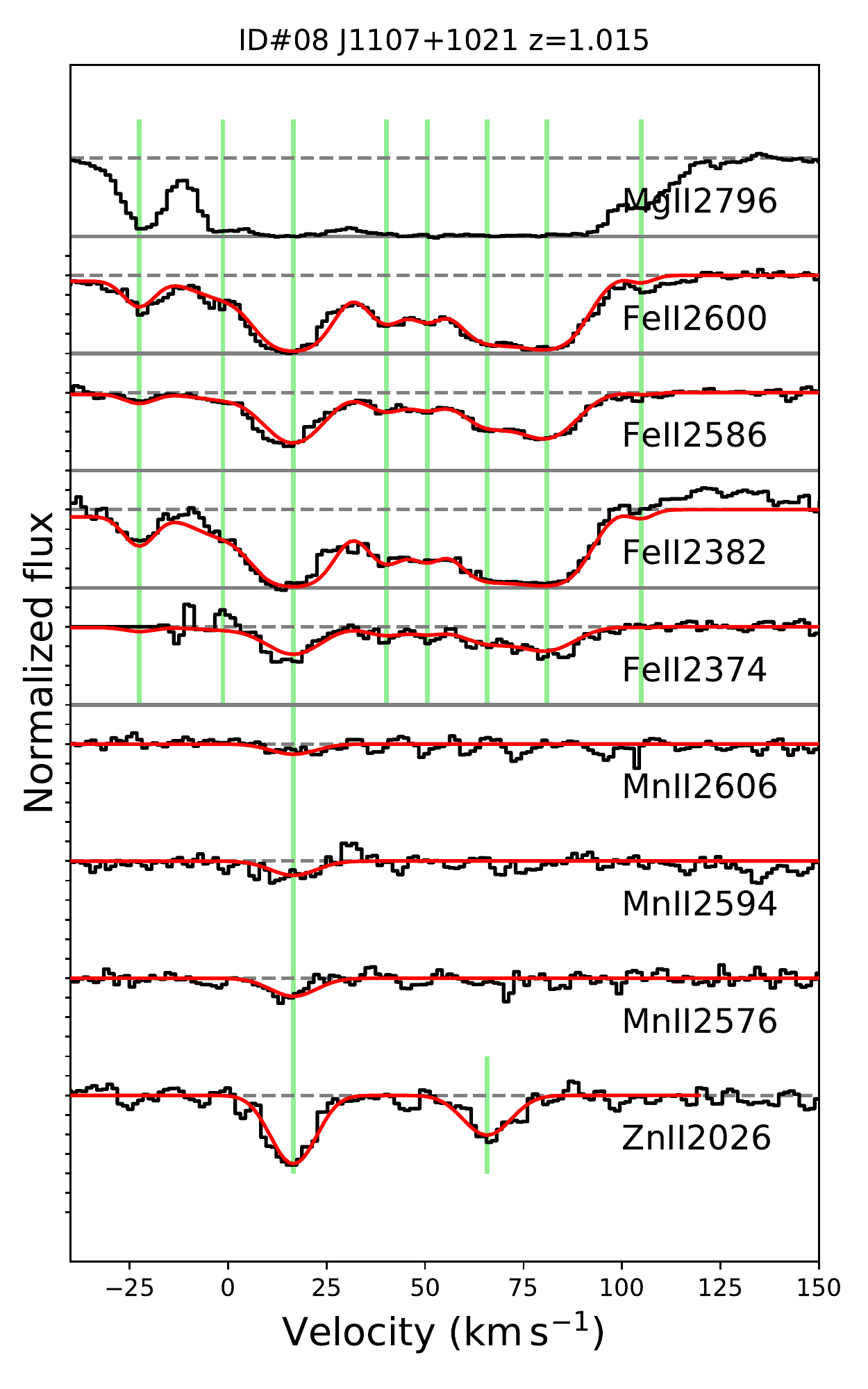}
\includegraphics[width=0.3\textwidth]{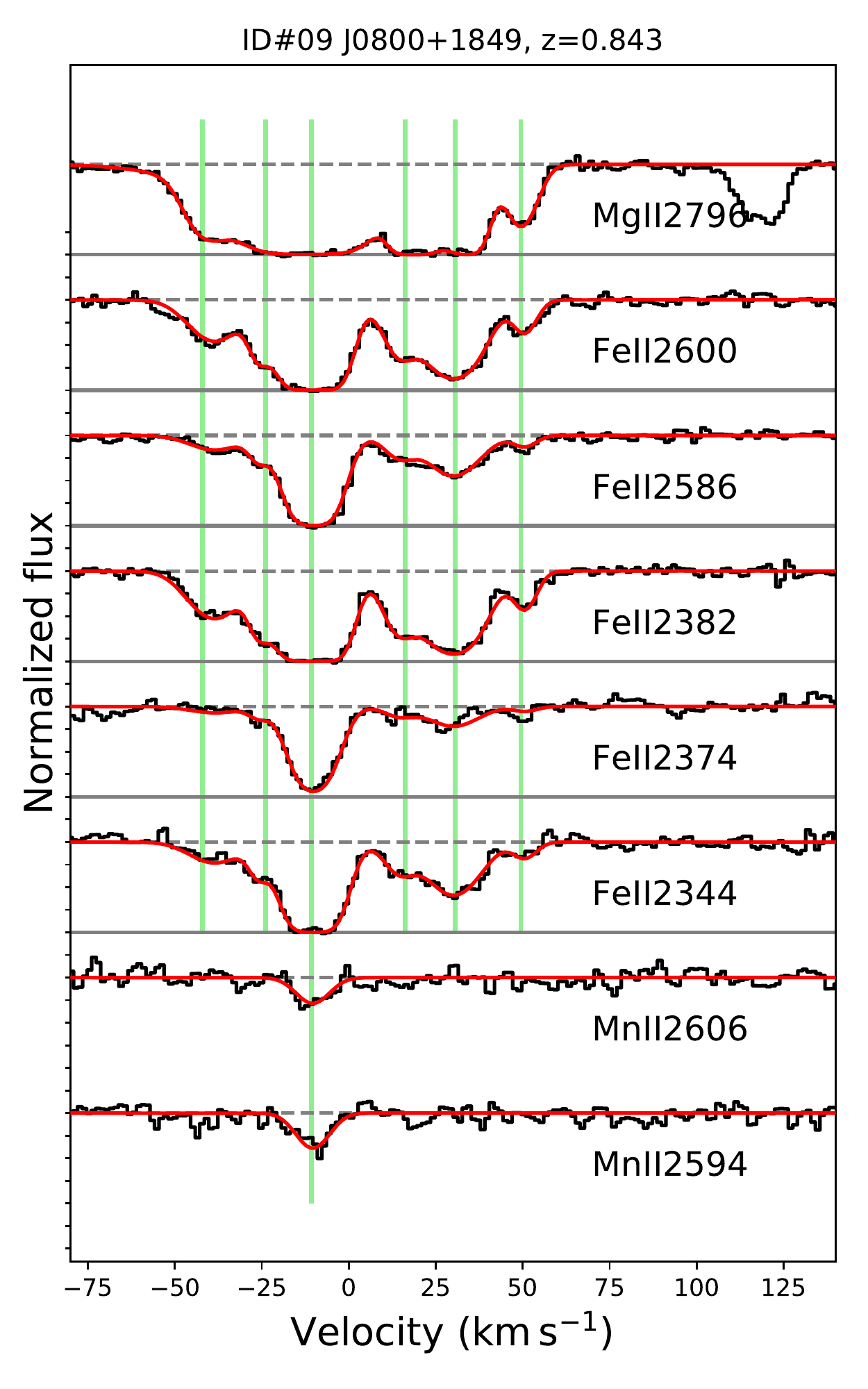}\\
\includegraphics[width=0.3\textwidth]{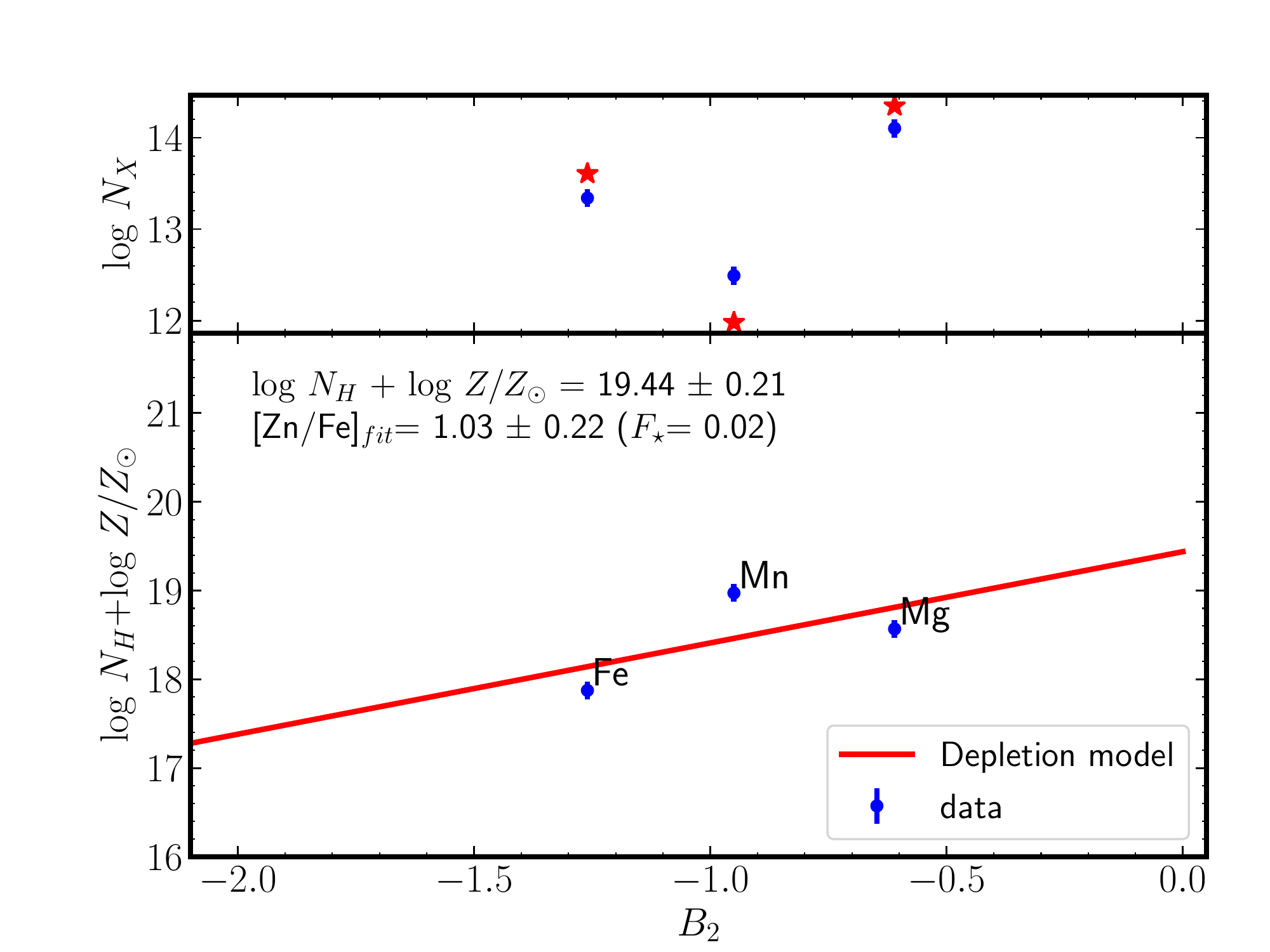}
\includegraphics[width=0.3\textwidth]{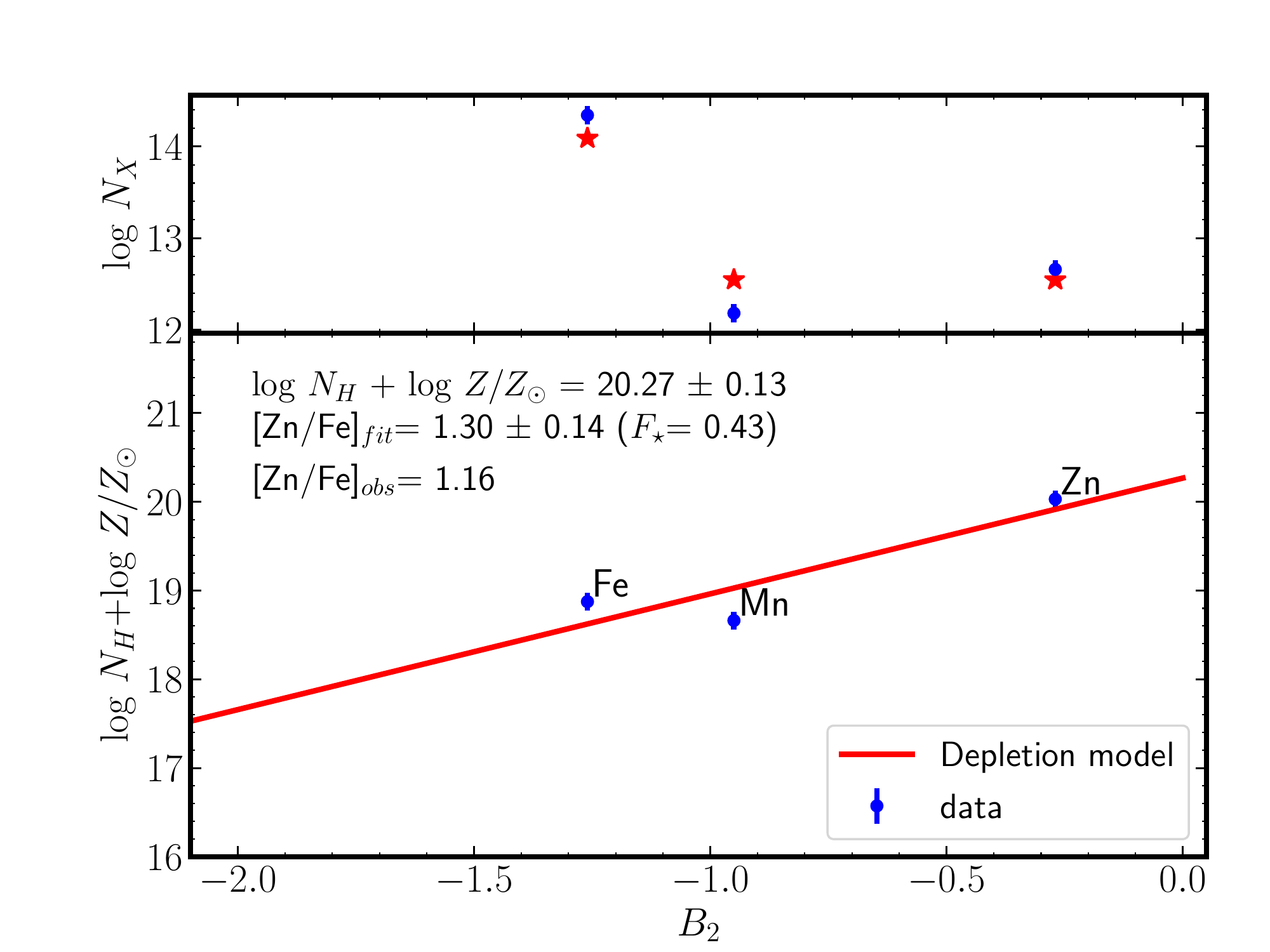}
\includegraphics[width=0.3\textwidth]{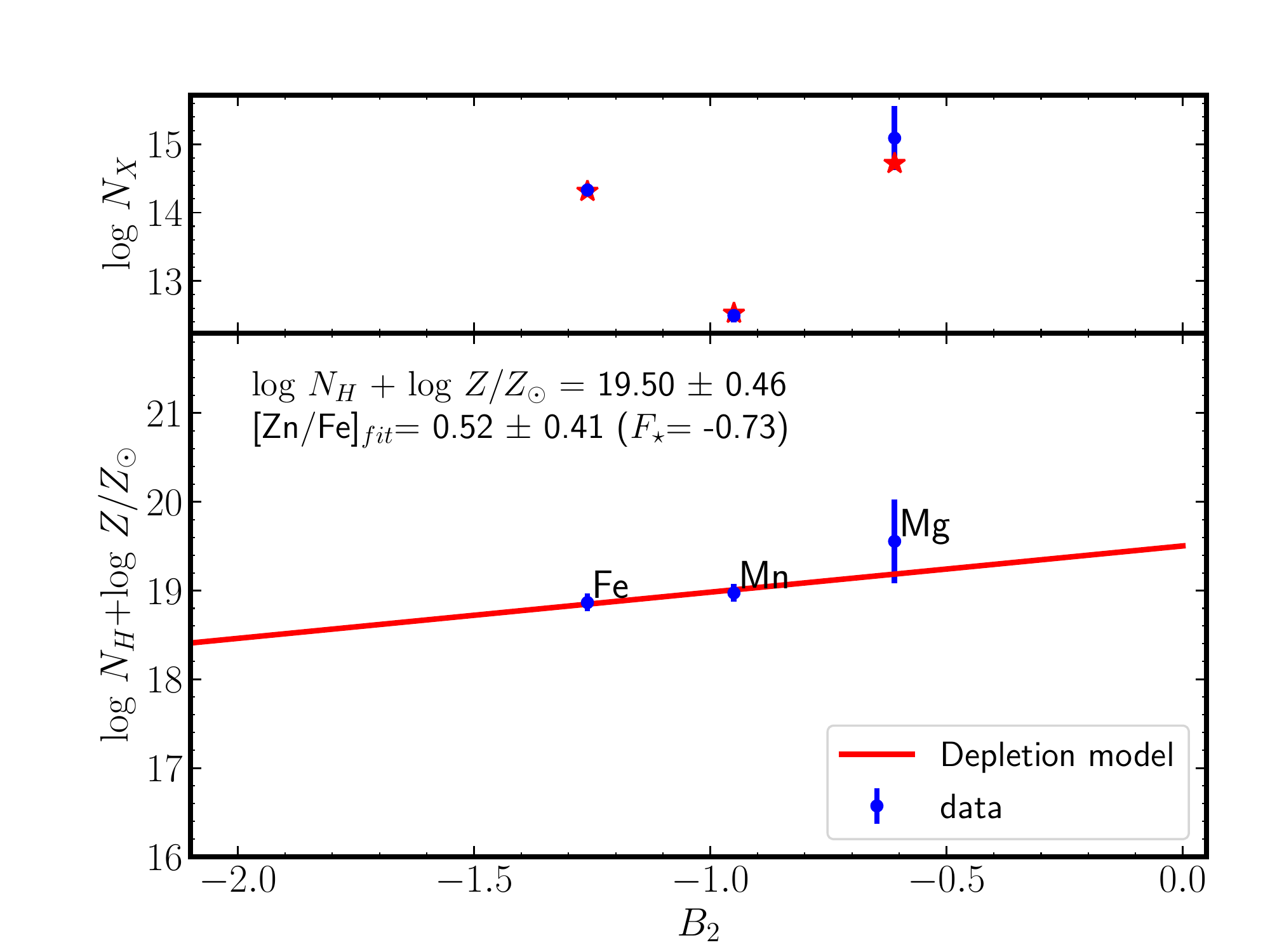}
\end{center}
\contcaption{Continued from Figure \ref{fig:depletion:examples}.}
\label{fig:example_cont2}
\end{figure*}
\begin{figure*}
\begin{center}
\includegraphics[width=0.3\textwidth]{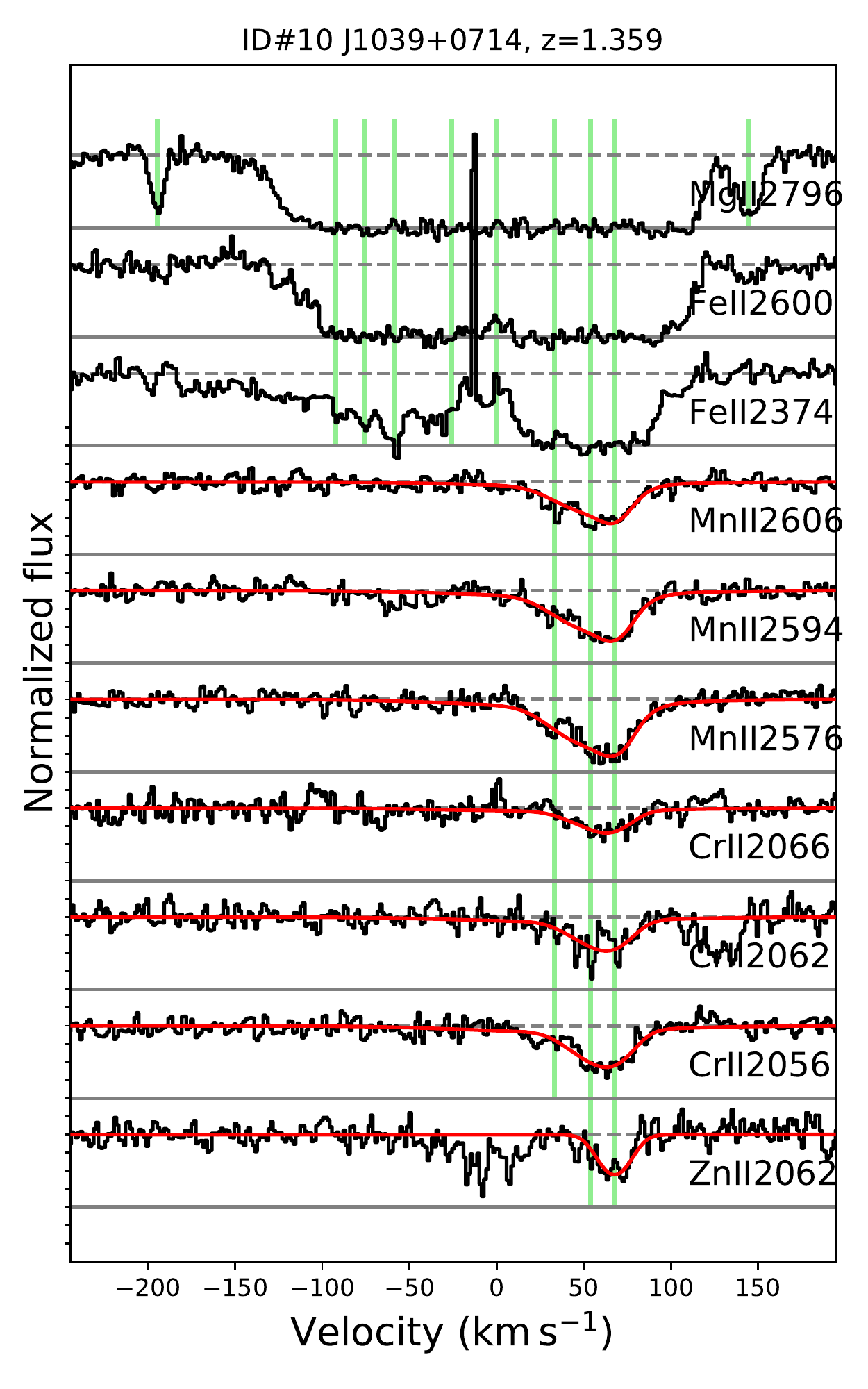}
\includegraphics[width=0.3\textwidth]{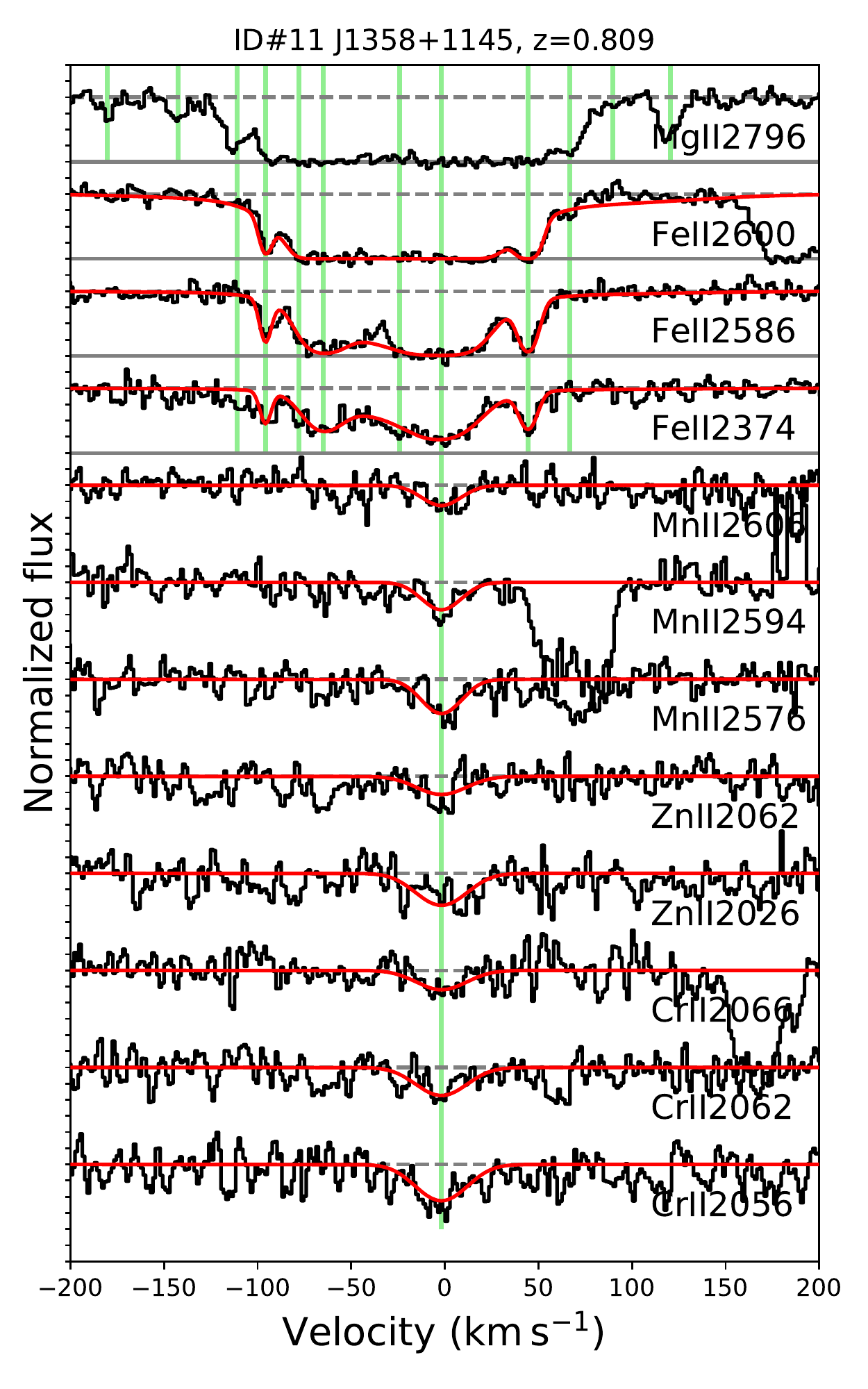}
\includegraphics[width=0.3\textwidth]{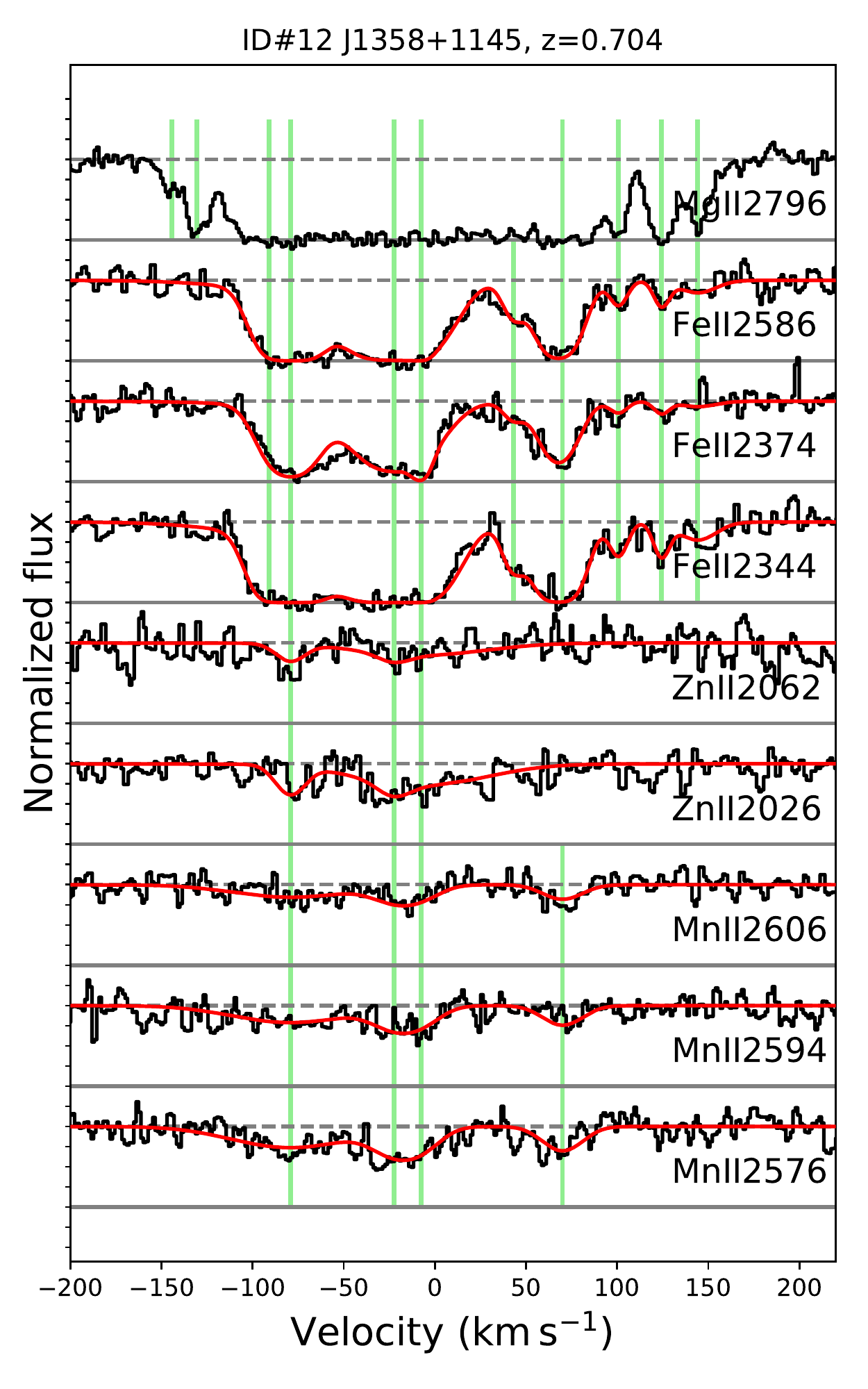}\\
\includegraphics[width=0.3\textwidth]{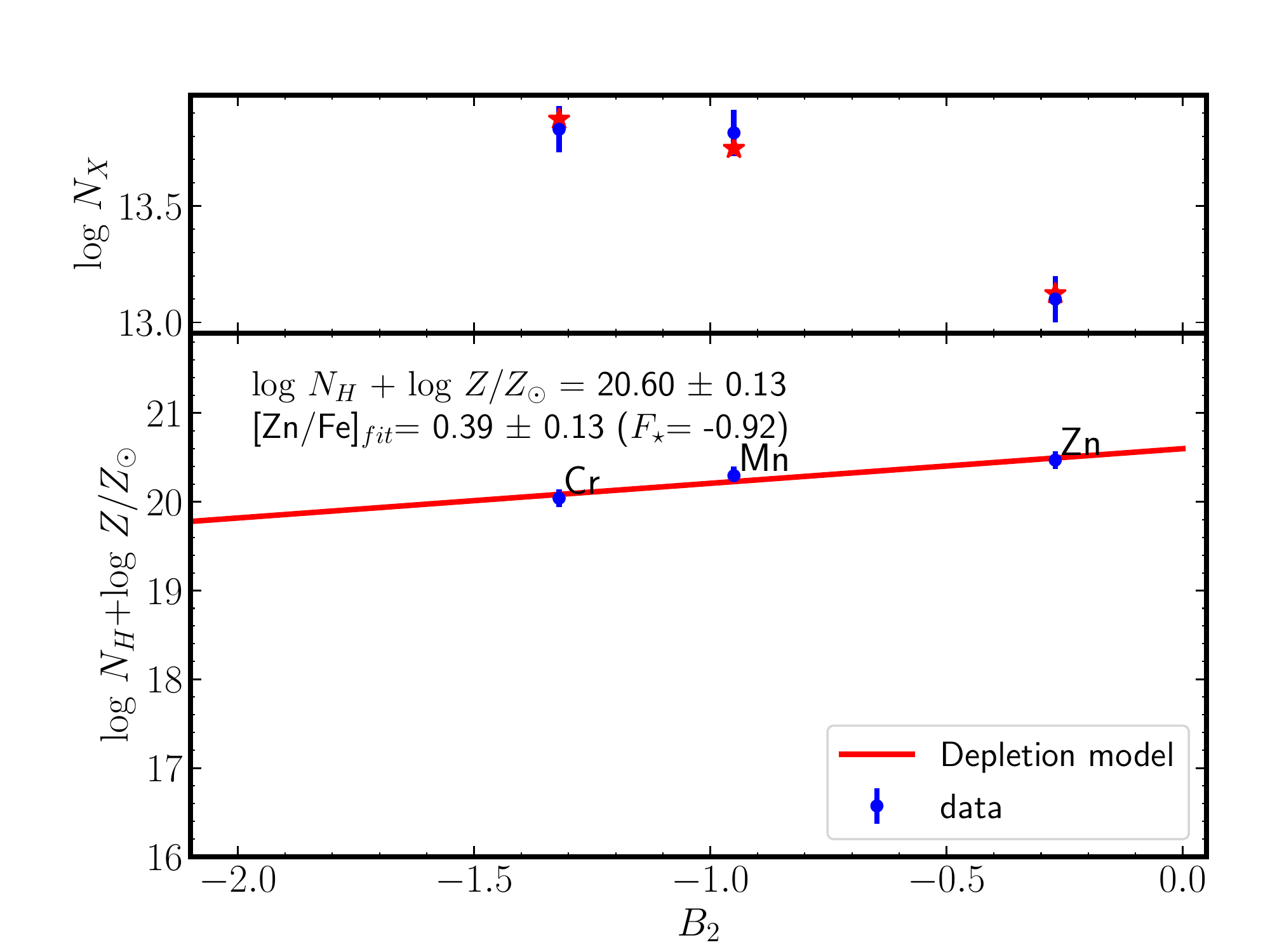}
\includegraphics[width=0.3\textwidth]{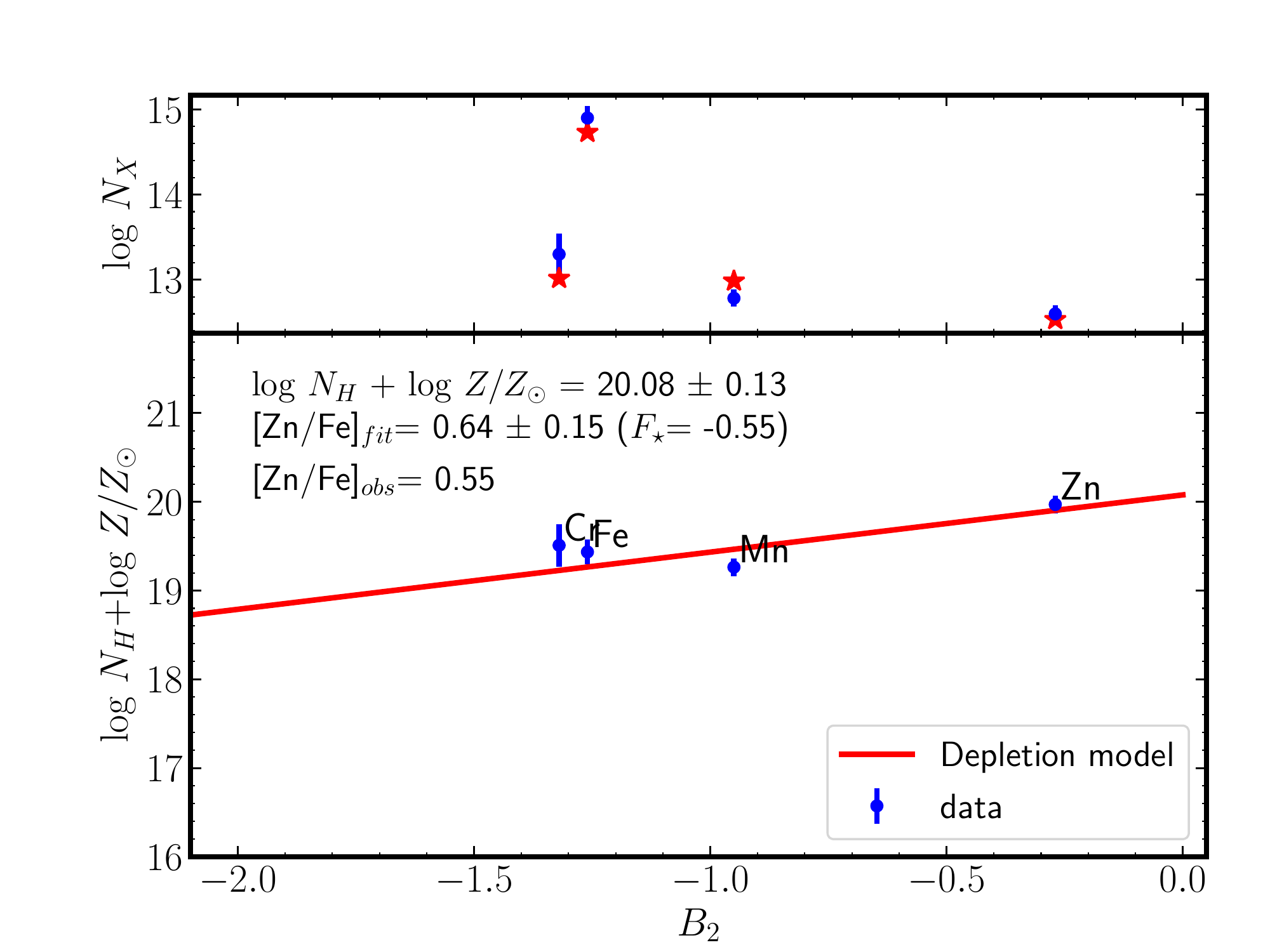}
\includegraphics[width=0.3\textwidth]{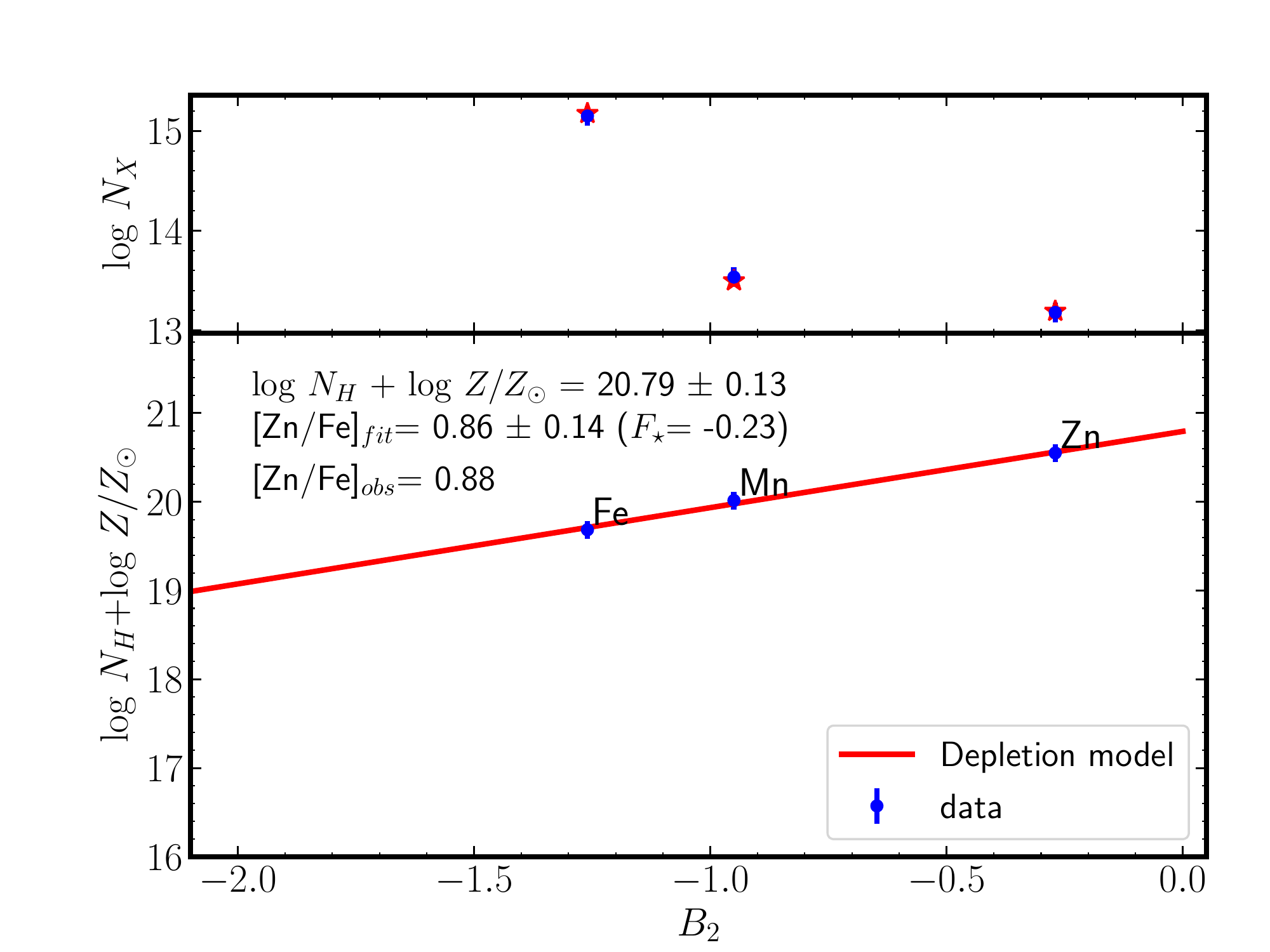}
\end{center}
\contcaption{Continued from Figure \ref{fig:depletion:examples}.}
\label{fig:example_cont3}
\end{figure*}
\begin{figure*}
\begin{center}
\includegraphics[width=0.3\textwidth]{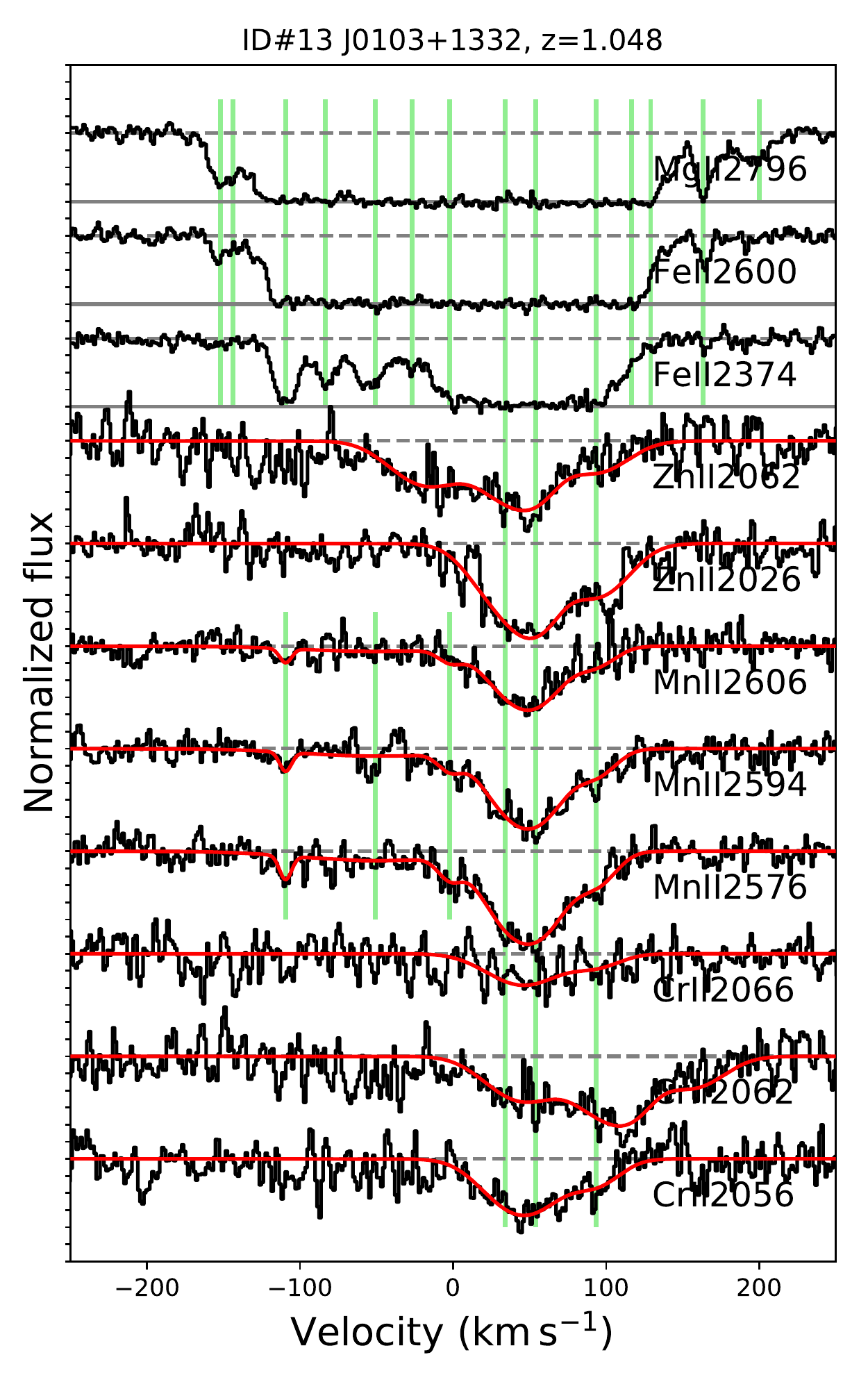}\\
\includegraphics[width=0.3\textwidth]{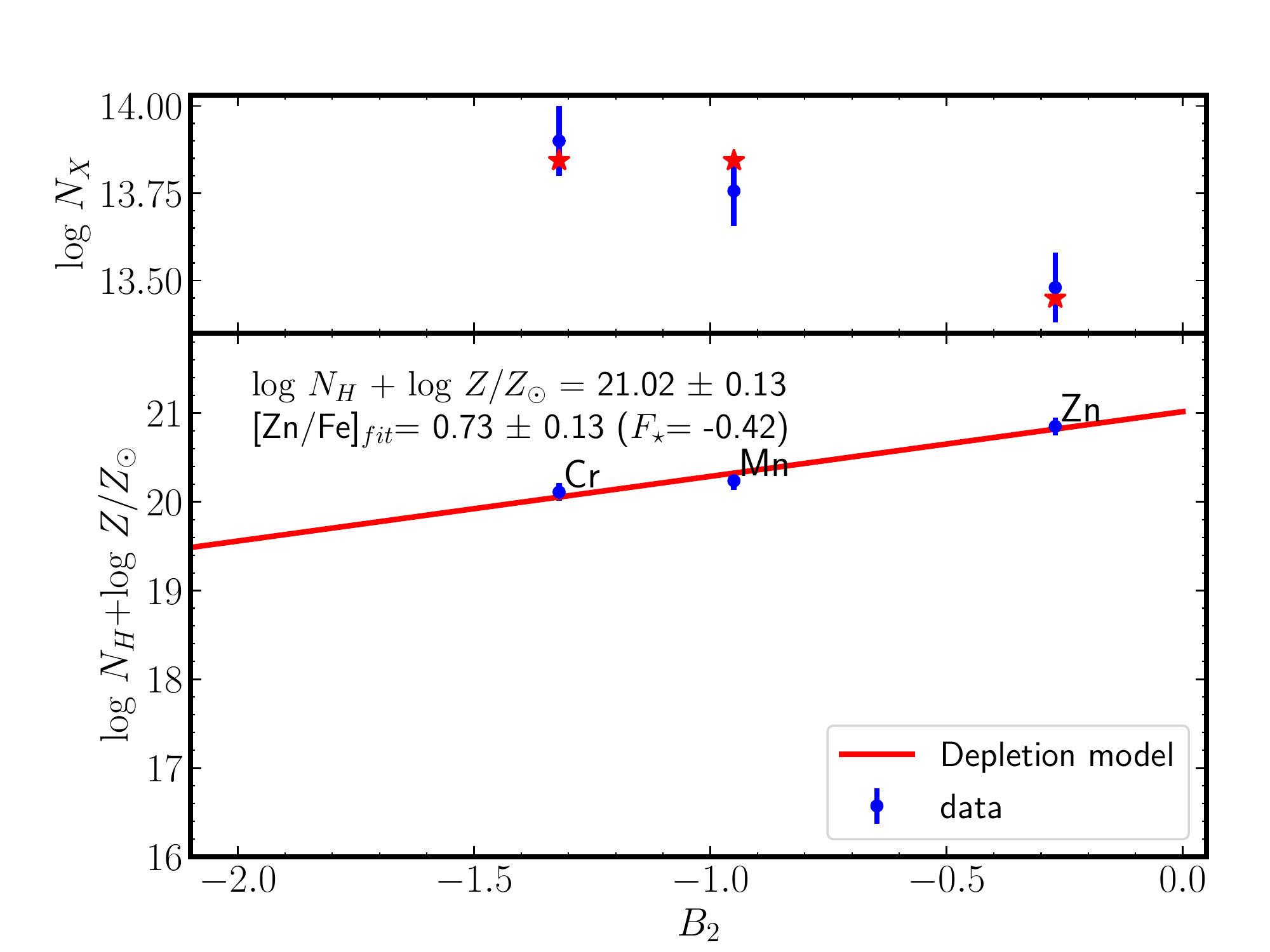}
\end{center}
\contcaption{Continued from Figure \ref{fig:depletion:examples}.}
\label{fig:example_cont4}
\end{figure*}
%
% Don't change these lines
\bsp	% typesetting comment
\label{lastpage}
\end{document}